\magnification=1200
\pageno=1

\centerline {\bf REMARKS ON THE EXISTENCE OF SPINNING MEMBRANE ACTIONS }
\centerline {\bf  }
\bigskip
\centerline { Carlos Castro}
\centerline { Center for Theoretical Sudies of Physical Systems}
\centerline { Clark Atlanta University, Atlanta, GA.30314}
\smallskip
\centerline {July , 2000 }
\bigskip
\bigskip

\centerline { \bf ABSTRACT}

It has been recently argued by some authors that is impossible to construct a Weyl invariant 
spinning membrane action, where the $S$-supersymmetry associated with the $3D$ superconformal algebra, 
is relinquished without gauge fixing. Contrary to those assertions, we show why it is possible to construct  
a Weyl-invariant spinning polynomial membrane action , without curvature terms, where $both$ the conformal 
boost symmetry and $S$-supersymmetry are explicitly broken by the action. It is shown that the gauge 
algebra $closes$ despite that the two latter symmetries are broken . 
For this to happen, a modifed $Q$-supersymmetry transformation , a sort of new $Q+K+S$ ``sum ``  rule, 
is required  that generates the compensating 
terms to cancel the spurious contributions from the $S$ and conformal boost anomalous 
transformations. A substantial discussion of  the quantization of the spinning membrane and anomalies 
is given. We review briefly the role that this spinning membrane action  may have in the theory of 
$D$-branes, Skyrmions and BPS monopoles in the large $N$-limit of $SU(N)$ Yang-Mills .

\bigskip

\centerline{\bf I. {INTRODUCTION}}
\bigskip

      In the past years there has been considerable
progress in the theory of extended objects, in particular two dimensional extended
objects ; i.e.  membranes. However, a satisfactory spinning
membrane Lagrangian has not been constructed yet, as far
as we know. Satisfactory in the sense that a suitable
action must be one which is polynomial in the fields,
without ( curvature) $R$  terms which interfere with the algebraic
elimination of the three-metric, and also where
supersymmetry is linearly realized in the space of
physical fields. It had been argued [1,2] that it was  allegedly 
impossible to
supersymmetrize Dirac-Nambu-Goto type of actions
(DNG)
-those proportional to the world-sheet and
world-volume spanned by the string (membrane) in their
motion through an embedding space-time. The efforts to
supersymmetrize this action have generally been based upon
the use of the standard, classically-equivalent, bosonic
action which included a cosmological constant. The
supposed obstruction is related to the fact that in order
to supersymmetrize this constant one had to include an 
Einstein-Hilbert term spoiling the process of the algebraic elimination of the three-metric altogether.

       Bergshoeff  et al [ 3] went evenfurther and
presented us with the "no-go" theorem for the spinning
membrane. Their finding was based in the study of a
family of actions, in addition to the one comprised of
the cosmological constant, which were equivalent, at the
classical level, to the DNG action. However, this "no-go"
theorem was not general enough.  because these authors relied on the
tensor calculus for Poincare D=3 N=1 SG developed by
Uematsu [ 4  ]. Unfortunately, the above tensor calculus despite being correct 
does $not$ yield a linearly realized supersymmetry for
the kinetic matter multiplet to start with. A constraint,
 $\bar{\chi}\chi$ =0  ,  appears after the elimination
of the $S$ auxiliary field, where $\chi$
is the $3D$ Majorna spinor. The spinning membrane requires supersymmetry on the world  volume 
whereas the supermembrane requires  target space-time supersymmetry. 
Lindstrom and Rocek [5]  were the first ones to
construct a Weyl invariant spinning membrane action.
However, such action was highly non-polynomial
complicating the quantization process evenfurther than the
one for the supermembrane and the membrane coordinates had a non-canonical dimension from the world volume 
point of view. 

     The suitable action to supersymmetrize is the one of
Dolan and Tchrakian [6] ( DT) without a cosmological constant
and with quadratic and quartic-derivative terms. 
Such membrane action is basically a Skyrmion  action with quartic and quadratic derivative terms. 
A class
of conformally-invariant $\sigma$- model actions was
shown to be equivalent, at the classical level, to the
DNG action for a p+1 extended object ( p+1=even) embedded
in a target spacetime of dimension $ d\geq {p+1}$. When
p+1=odd, our case, an equivalent action was also
constructed, however, conformal invariance was lost in
this case. The crux  of the work presented here  lies in the necessity to embed the Dolan-Tchrakian action into 
an explicitly Weyl invariant one through the introduction of extra fields. These are the gauge field of 
dilations , $b_\mu$, not to be confused with the $U(1)$ world-volume gauge field appearing in $D$-branes, and 
the scalar coupling , $A_0$ of dimension $(length)^3$, that must appear in front of the quartic 
derivative terms of the DT action. The latter terms must appear with a suitable coupling constant in order to 
render the action dimensionless. As a result of the embedding into a Weyl invariant action the coupling  $A_0$ 
turns into a scalar of Weyl weight equal to $-3$.

Having Weyl-covariantized the DT action, the natural question to ask is how do we eliminate these
new fields, $b_\mu, A_0$ in order to recover the original action ? This can be achieved simultaneously 
if one first notices that upon the Weyl covariantization process , 
the multiplicative coupling {\bf constant}  
$g$, obying $\partial_\mu g =0$,  is $promoted$ 
to a scalar-like parameter $A_o$ obeying  the Weyl covariant 
extension : $D^{Weyl}_\mu A_0 =0$. 
Since $A_o$ has a nontrivial Weyl weight, the  latter condition 
implies that $b_\mu \sim \partial_\mu ln(A_o)$.  A condition on $b_\mu$ such 
that upon gauging the scalar-like $A_o$ back to a constant $g$,  
yields $b_\mu = 0$ . 
It would be meaningless to vary those multiplicative couplings, like $A_o$, 
in the action because this variation will constrain the action to zero. 
Such couplings are $not$  Lagrange multipliers to be varied. One does not vary multiplicative couplings in the actions. 

One must concentrate solely on the Weyl covariantized version of the Dolan-Tchrakian action, because $it~is$ the Weyl-covariantized DT action that 
we will be supersymmetrizing and $not$ the orginal DT action. 
One will obtain in this manner an  action that is invariant under 
a supersymmetric extension of the homothecy transformations.     

For these reasons,  we must emphasize that one is {\bf not} 
imposing {\bf by hand} a capricious constraint on the multiplicative coupling $A_0$. On the contray, Weyl covariance {\bf demands } the natural 
extension of the condition $\partial_\mu g =0$ to the Weyl covariant version $D^{Weyl}_\mu A_0 =0$. 
Hence, the original DT action is attained simply by 
fixing the Weyl gauge invariance upon setting $A_o =g$.  This gauge choice in conjunction with the embedding condition, $D^{Weyl}_\mu A_0 =0$, that is equivalent to the pure gauge condition:  $b_u \sim \partial_\mu lnA_0$ , will finally render the gauge field $b_\mu =0$, furnishing the original DT action as 
expected.  

The above condition, $D^{Weyl}_\mu A_0 =0$,       
which simply states the fact that the coupling $A_0$ 
is a " constant " from the Weyl covariant point of view, is what we call the " embedding condition" of the DT action into the Weyl-covariantized Dolan-Tchrakian version (WCDT). 
There  might be global topological obstructions to set $D_\mu^{Weyl} A_o =0$ globally; i.e having a pure gauge $b_\mu\sim \partial_\mu ln (A_o)$ everywhere,  that are not discussed here [7] which may be very relevant in the nonperturbative behaviour of the theory.  

To support the fact that the embedding condition is a natural consequence of the Weyl covariance/invariance of the Lagrangian, we will also show ( in Appendix {\bf A}) 
that the condition  $D^{Weyl}_\mu A_0 =0$ is also $compatible $  with the equations of motion of the membrane's physical matter fields.
It follows that if the equations of motion of the matter ( physical) membrane fields associated with  the Weyl covariantized  DT action (WCDT)  are indeed the
Weyl covariant extension of the original DT equations of motion for the matter fields, then the condition $D^{Weyl}_\mu A_0 =0$  is indeed consistent; i.e compatible.  

To sum up, the embedding condition    $D^{Weyl}_\mu A_0 =0$  is 
tantamount of saying that the coupling scalar $A_0$ is just the analog of a constant from the Weyl covariant point of view : it is  
the Weyl covariant extension of the original ( not Weyl covariant) condition on the  coupling  constant, $g$ : $\partial_\mu g = 0$. Therefore, from the 
WCDT  action we may  recover the original DT action upon choosing the gauge $A_0=g$ and implementing  the embedding condition 
 $D^{Weyl}_\mu A_0 =0$. The superconformaly covariant extension of the WCDT action requires to use then the corresponding superconformaly covariant constraint 
$D^C_\mu A_0 =0$ where the superconformal 
covariant derivative is given in the text.

Once the embedding program  into the WCDT action has been performed one supersymmetrizes the WCDT action by 
incorporating $A_0$ into the superconformal coupling-function multiplet $(A_0, \chi_0, F_0)$;  whereas the $b_\mu$
becomes part of the gravitational conformal supermultiplet involving $(e^m_\mu, \psi_\mu, b_\mu)$ and the physical matter fields 
of the membrane form part of the world-volume superconformal multiplet $(A^i, \chi^i, F^i)$. 
The $A^i$ fields are identified with the membrane target space time coordinates. 
It is shown in this work that the background gravitational supermultiplet and the coupling supermultiplet are fully 
determined in terms of the membrane physical fields after using the embedding condition  $D^c_\mu A_0 =0$, 
and after the elimination of the members of the gravitational superconformal multiplet via their nonpropagating equations of motion.  
Thus there is $no$ need to vary the fields $F_0, A_0,\chi_0$ in order to eliminate them algebraically as we did in [8]. It is sufficient to vary the gravitational supermultiplet supplemented by the superconformal covariant embedding condition  
 $D^c_\mu A_0 =0$. This is one of the main new points we wished to add to the spinning membrane literature.

If one wishes to eliminate any curvature scalar terms in the final action  one must take 
suitable combinations of these three superconformal multiplets  and, in doing so, one is going to break  
$explicitly$ the $S$ supersymmetry of the $3D$ superconformal algebra as well as the conformal boost symmetry, the $K$ symmetry,  
which signals the
$presence $ of the $b_\mu$ field in the final action : it does $not$ decouple as it does in ordinary 
$3D,4D$ superconformal gravitational actions where the $full$ superconformal invariance is maintained that 
allows to fix the conformal boost $K$-symmetry by choosing the gauge condition  $b_\mu =0$.

 The final action is Lorentz, dilational, ( a modified ) ${\cal Q} $ supersymmetric and translational invariant but is $not$ 
invariant under $S$ supersymmetry nor conformal boosts, $K$ . In principle, there is nothing 
$wrong$ with this fact because the rigid subalgebra  of the full $3D$ superconformal algebra comprised of the Lorentz
generator, dilations, $Q$ supersymmetry and translations, $P_\mu$, does $close$!. 
It remains to prove that the $local$  algebra also $closes$. For this to happen, a modified ${\cal Q}$-supersymmetry variation will be constructed which generates those compensating terms that cancel the spurious contributions stemming from the $S$ and $K$ breaking anomalous transformations.

We will prove in {\bf 3.2} that the commutator of two modified ${\tilde \delta}_Q$ transformations does close. In conventional Poincare supergravity one has invariance only under a particular combination of $Q$ and $S$ supersymmetry, the 
so called  $Q+S$ sum rule and the original $K$ symmetry is used to  enforce the decoupling of the $b_\mu$ from the action  by fixing the conformal boost symmetry  $b_\mu =0$. 
Here we have a different picture, we have a modified  ${\cal Q} $ supersymmetry, which can be thought of a $Q+K+S$ sum rule, instead of a particular combination of $Q$ and $S$ which occurs in Poincare Supergravity, 
and there is  $no$ explicit conformal boost invariance to start with. 
Hence, there is no need to relinquish the $S$ nor conformal boosts, by  gauge fixing such symmteries 
$by~ hand$ ,  because the action itself breaks both the $S$-supersymmetry and conformal boost symmetry 
explicitly.

The outline of this work goes as follows. In the first part of section II we present the work of Dolan and
Tchrakian and show how upon the algebraic elimination of the auxiliary world-volume metric from the action, one recovers the Dirac-Nambu-Goto action for the membrane. 
A Weyl covariantization 
program of the DT action, and its subsequent algebraic elimination of the world-volume metric,  furnishes the Weyl covariant extension of the DNG action with the provision 
that the membrane Tension, 
is in this Weyl covariant/invariant  case, a suitable scalar-like parameter whose nontrivial Weyl weight has a negative value with respect to the Weyl weight of the 
multiplicative scalar coupling $A_o$ associated with the quartic derivative terms. 
This is precisely what is needed for the Weyl transformation of the Tension to compensate for the nontrivial Weyl transformation of the world-volume of the membrane 
associated with the DNG action. One has then a Weyl invariant membrane.  
In the string case everything is explicitly Weyl invariant to begin with, so this procedure is automatic. The spinning string is superconformally invariant. 

A further discussion of  the problems
associated with the $3D$ Poincare Supergravity tensor calculus is presented :  
a constraint arises among the physical fields 
upon elimination of the auxiliary fields. Therefore, this tensor calculus is not suited to have a satisfactory spinning membrane action.  
In the last part of {\bf II} the connection between the supersymmetric version of the generalized Skyrme's   model of baryons  
and  the   
spinning membrane propagating in a curved backgrounds is studied and the role that the 
spinning membrane may have in the physics of D-branes and BPS monopoles is discussed. 
It is suggested that $D0$-branes, in the large $N$ limit ,  may play an important role in the spectrum of the generalized Skyrmion action .    

In {\bf III} , we provide   the detailed  arguments showing that in
order to satisfy all of the stringent requirements
discussed earlier in order to have a satisfactory spinning membrane we must relinquish $S$ 
supersymmetry $and$ conformal boosts and  
concentrate solely on the ( modified)  ${\cal Q}$-supersymmetry associated with
the superconformal algebra in three dimensions. The Poincare $"Q+S"$ 
supersymmetry can only be implemented in the class of
non-polynomial actions (if we insist in meeting all of our
requirements) as it was shown in  [5] . The fully
${\cal Q}$-invariant action is furnished providing  a ${\cal Q}$-spinning  and Weyl invariant  polynomial 
membrane action without curvature terms ( nor gravitino kinetic terms ) . 

In section {\bf 3.1} we present the supersymmetrization of the 
quadratic  kinetic terms and discuss  how the gauge algebra $closes$ after introducing the modifed 
${\cal Q}$-transformations to cancel out the spurious terms stemming from the anomalous $S, K$ variations.  
In section {\bf 3.2} we show the closure of the gauge algebra and add some important remarks. 
In {\bf 3.3} the supersymmetrization of the quartic  derivative terms is presented.   
Finally, in {\bf 3.4} we show that  upon the elimination of the 
auxiliary fields $F^i$ no constraints arise among the membrane fields.  
In {\bf IV} we give a discussion of the quantization program of the spinning membrane action  and how the absence of anomalies of the light-cone Lorentz algebra algebra should select 
$D=11$ as the critical spacetime embedding dimension.

To finalize, in appendix {\bf A} the embedding condition  $D^c_\mu A_0 =0$ is shown to be compatible ( consistent)  with the equations of motion of the membrane matter fields associated with the WCDT action. 
The latter condition is essential in order to retrieve the Dolan-Tchrakian action after the elimination of theauxiliary $F^i$ fields, setting the fermions to zero and fixing, finally,  the gauge $A_0 =g$.

Our coventions are: Greek indices stand for
three-dimensional ones; Latin indices for spacetime ones : $i,j =0,1,2....D$. The signature of the $3D$ volume is $(-,+,+)$. 

\bigskip
\centerline{\bf II}
\smallskip

\centerline {\bf 2.1 The Dolan-Tchrakian Action}
\bigskip

The Lagrangian for the bosonic $p$-brane ( extendon) with vanishing cosmological constant constructed by Dolan and
Tchrakian in the case that $p=odd; p+1=2n$ is :

$$L_{2n} =\sqrt {-g} g^{\mu_1 \nu_1}.... g^{\mu_n \nu_n} \partial_{[\mu_1} X^{i_1}.........\partial_{\mu_n]} X^{i_n}
\partial_{[\nu_1} X^{j_1}.........\partial_{\nu_n]} X^{j_n} \eta_{i_1 j_1}.... \eta_{i_n j_n}. \eqno (2-1)$$
$\eta_{ij}$ is the spacetime metric and $g^{\mu\nu}$ is the world volume metric of the $2n$ hypersurface spanned by the motion
of the $p$-brane. Antisymmetrization of all indices is also required and because of this, the DT action is at most quadratic in time derivatives which does 
not spoil the quantization process. 
The action is also conformally invariant in 
even worldvolume dimensions.

Upon elimination of the world volume metric encoded in the   
$2n\times 2n$ matrix :

$$ A^\mu_\nu =g^{\mu \rho} \partial_{\rho} X^{i}(\sigma)\partial_{\nu} X^{j}(\sigma)\eta_{ij}. \eqno (2-2) $$

via its equations of motion : $ { \delta L_{2n} \over \delta A} = 0 
\Rightarrow $ gives the $n^{th}$ order matrix polynomial equation 
in $A^\mu_\nu$:

$$ A^n -b_{2n-1}A^{n-1} +b_{2n-2}A^{n-2}..................(-1)^{n+1} b_{n+1}A+{1\over 2} (-1)^n b_n I_{2n\times 2n}=0. \eqno (2-3) $$
where  the scalar coefficients appearing in the matrix-polynomial equation (2-3) can be read from the first $n+1$ coefficients in 
the expansion of :

$$det(A^\mu_\nu -\lambda I_{2n\times 2n}) =\lambda^{2n} -b_{2n-1} \lambda^{2n-1}+b_{2n-2} 
\lambda^{2n-2}..........-b_1\lambda +det(A^\mu_\nu). \eqno (2-4)$$

After substituting  the matrix solution to the matrix-polynomial equation (2-3) with coefficients given by (2-4) back into the action (2-1) yields :

$$L_{DNG} =\sqrt {-det( \partial_{\mu} X^{i}(\sigma)\partial_{\nu} X^{j}(\sigma)\eta_{ij})}~[{(n!)^2 b_n \over \sqrt A}]. \eqno (2-5) $$
The crucial observation made by [6] is that the last factor :

$$[{(n!)^2 b_n  \over \sqrt {A}}].~~~A=det~A^\mu_\nu. \eqno (2-6)$$
takes discrete $numerical$   values for all values of $n$. 
Therefore , the equivalence to the Dirac-Nambu-Goto action has been established. Notice that for every $n$,  the Dolan-Tchrakian Lagrangian $L_{2n}$ is conformal invariant and it is only quadratic in time derivatives due to the antisymmetry of the indices.  This, however, does not imply that the Dirac-Nambu-Goto action is quadratic in time derivatives . Since $L_{2n}$ is polynomial in the fields, attempts to quantization might not be hopeless.  

When $p=even$ , $p+1$ odd,  a Lagrangian with zero cosmological constant can also be constructed, however, conformal invariance is $lost$.
In the membrane's case one has :

$$L=L_4 +aL_2.~~~a>0. ~~~L_2 =\sqrt {-g} g^{\mu\nu}\partial_{\mu} X^{i}(\sigma)\partial_{\nu} X^{j}(\sigma)\eta_{ij}. $$
$$L_4 =\sqrt {-g} g^{\mu\nu}  g^{\rho \tau}\partial_{[\mu} X^{i}(\sigma)\partial_{\rho ]} X^{k}(\sigma)\partial_{[\nu} X^{j}(\sigma)\partial_{\tau ]} X^{l}(\sigma)
\eta_{ij} \eta_{kl}             . \eqno (2-7) $$

Upon elimination of the world volume metric in ( 2-7) and plugging its value back into the action gives the Dirac-Nambu-Goto action  : 

$$12 \sqrt {a} \sqrt {-det ~G_{\mu\nu}}~or~  -4\sqrt {a} \sqrt {-det ~G_{\mu\nu}}. \eqno (2-8)$$
with the induced worldvlume metric being :  

$$G_{\mu\nu}= \partial_{\mu} X^{i}(\sigma)\partial_{\nu} X^{j}(\sigma)\eta_{ij}. \eqno (2-9) $$
Notice that $a >0$ in (2-7) so both $L_2$ and $L_4$ have the same relative sign.  

When a Weyl covariantization program of the DT action is taken, the elimination of the worldvolume metric yields the Weyl covariant version of the DNG action ( where now we are explictitly 
including the Tension in our expressions ) :

$$S \sim \int d^3\sigma T \sqrt a  \sqrt {-det~ G_{\mu\nu}}. ~~~G_{\mu\nu }= A_o^{{1\over 3}}  (D^{Weyl} _\mu X^i)( D^{Weyl} _\nu X^j) \eta_{ij}. \eqno (2-10)$$ 
we have written the induced three metric, as a result of the embedding , up to a {\bf crucial}  conformal factor $A_o^{1/3}$  which precisely contains the coupling $A_o$ as it should, to ensure that the action is truly Weyl covariant.   
The factor $a$ in eqs-(2-7, 2-8) has for  numerical value the ratio of the ( numerical ) quadratic couplings to the ( numerical ) quartic couplings : 
$a/1 =g_2/g_4$. Setting the numerical value $g_4 =A_o $ implies that the $numerical$ value  for the quadratic coefficient is $g_2=a A_o$. It is this factor 
of $\sqrt g_2 = \sqrt {a A_o }$ that must appear in the Weyl covariantized Dirac-Nambu-Goto action (2-10). 
One can reshuffle the $\sqrt A_o$ factor inside the ${\sqrt G_{\mu\nu}} $ terms as we did explicitly in eq-(2-10).
 For this reason, in $D=3$ there is the explicit conformal factor $A_o^{1/3}$ multiplying the Weyl covariant derivatives of the membrane coordinates in eq-(2-10). 
We must have this prefactor $A_o^{1/3} $ in order to ensure that the action is Weyl invariant.    
This occurs in the cases that conformal invariance no longer holds : the cases when $p+1 =odd$. In the membrane case the Weyl scaling behaviour of the 
induced three metric $g_{\mu\nu}$ , given by 
the pullback of the target spacetime metric 
$\eta_{ij}$ to the worldvolume of the membrane,  is :

$$  g_{\mu\nu}= (D^{Weyl} _\mu X^i) (D^{Weyl} _\nu X^j) \eta_{ij} \rightarrow  e^{-\lambda} g_{\mu\nu} =  e^{-\lambda}  (D^{Weyl} _\mu X^i)
( D^{Weyl} _\nu X^j) \eta_{ij}. \eqno (2-11) $$
due to the fact that the Weyl weights of the membrane $X^i$ coordinates are {\bf no }  longer $0$ (like in the conformally invariant cases). 
These weights are given instead by ${1\over 2}$, in units of $\lambda$. As a result, the induced three metric , $g_{\mu\nu}$, does {\bf not} have
the appropriate Weyl weight of $-2\lambda$ as it is required. The factor of $A_o^{1/3}$ in eq-(2-10) is crucial 
since it has precisely the right compensating Weyl weight of $-\lambda$ to ensure Weyl invariance of the action. 
The  Weyl weight assignement of $A_o$ was $-3$ ( in units of $\lambda$)  consistent with its dimensions of $(length)^3$. . 
 
Concluding, with the inclusion of the right factor of $A_o^{1/3}$, the rescaled three metric : $G_{\mu\nu} = A_o^{1/3} g_{\mu\nu}$ 
scales as expected : with the suitable $e^{-2\lambda} $ factor. The Membrane Tension has 
the dimensions of $length^{-3} $ , and consequently,  a Weyl weight of negative sign 
relative to that of $A_o$ , 
i.e : weight of Tension is $3$. This is what is required for  the action eq-(2-10) to be invariant under Weyl transformations. 
The Weyl transformation of the Tension scalar-like parameter compensates with the Weyl scaling of the world volume ( square root terms). 
Finally,  the action is Weyl invariant under the following scalings :  

$$T \rightarrow e^{3\lambda} T.~~~ \sqrt {-det G_{\mu\nu}} \rightarrow e^{-3\lambda}\sqrt {-det G_{\mu\nu}}. ~~~ G_{\mu\nu} \rightarrow e^{-2\lambda} G_{\mu\nu}. 
\eqno (2-11)$$   
This intermediate step ( Weyl covariantization ) is unnecessary in the string and $p+1 =even $ cases, since the bosonic actions  are  already conformally invariant to start with. 
The Weyl weights of the string coordinates, and $p$-brane coordinates when $p+1 =even$ are 
$0$ since their dimensions are also $0$. Their DT actions are automatically conformally invariant in even world volume dimensions $D=2n$.

Concluding this section, we have shown how the DT action , upon elimination of the worldvolume metric, yields the Dirac-Nambu-Goto actions and shown the importance of the $A_o$ scalar-like parameter to implement Weyl covariance 
of the Weyl covariant extension of the DT membrane action before it is suprsymetrized.

\smallskip

\centerline {\bf 2.2 A nonlinearly realized Poincare-supersymmetric Membrane}
\bigskip

The Poincare-supersymmetric kinetic terms ( modulo total derivatives) for the $3D$ Poincare supergravity 
was given by Uematsu [4]. The fields are $\Sigma_P=(A,\chi, F')$ where one must not confuse the 
auxiliary field $F'$ with the one of the superconformal multiplet : $F=F'+{1\over 4} AS$. 
An invariant action can be constructed using the tensor calculus [4] from the supermultiplet $\Sigma_P$ and its 
kinetic scalar multiplet $T_P(\Sigma_P)$ as follows  :

$$L={1\over 2} [\Sigma_P \otimes T(\Sigma_P)]_{inv} -{1\over 4} [T(\Sigma_P \otimes \Sigma_P)]_{inv}=$$

$$e[-{1\over 2} g^{\mu\nu} \partial_\mu A \partial_\nu A -{1\over 2} {\bar \chi} \gamma^\mu D_\mu \chi +
{1\over 2} F'^2 +{1\over 2}  {\bar \psi}_\nu \gamma^\mu \gamma^\nu \chi \partial _\mu A +$$
$$
{1\over 16} {\bar \chi}\chi {\bar \psi}_\nu \gamma^\mu \gamma^\nu\psi_\mu +{1\over 8} S{\bar \chi}   \chi].
\eqno ( 2-12)$$
The Lagrangian is essentially identical to the Neveu-Ramond-Schwarz spinning string with the crucial difference that 
the ``effective mass'' term $S{\bar \chi} \chi$ is $not$ present in the string case ! 
Therefore upon eliminating the auxiliary field $S$ in the action (2-12) yields the unwanted constraint :  
$S{\bar \chi} \chi=0$ that spoils the linear realization of supersymmetry to start with ! In order to remedy 
this one could add the pure supergravity action with a corresponding $S^2$ term; however, this is precisely
what one wanted to avoid : the presence of $R$ terms in our action. 
Despite being able to write down a Poincare-supersymmetric extension of the DT action one still will be 
faced with the problem that supersymmetry will not be linearly realized upon elimination of the 
$S$ auxiliary fields.

There are ways to circumvent this 
problem. One way was achieved by Linstrom and Rocek who started with a non-polynomial  
Weyl invariant action (  if the membrane coordinates had a non-canonical Weyl weight of zero )  :

$$I\sim \int d^3\sigma~\sqrt {-g} [g^{\mu\nu}\partial_\mu X^i \partial _\nu X^j \eta_{ij}]^{3/2}.\eqno(2-13)$$
Such nonstandard actions that violated Derrick's theorem like  (2-13) were  considered earlier by Duff et al [11] in the study of kinks in $3D$. 
Since the auxiliary field $S$ is an alien concept in conformal supergravity it cannot appear  in the supersymmetrization process
where one uses conformal supergravity techniques to build invariant actions. Notice the $noncanonical$  dimension of the coordinates in 
(2-13) $X^\mu$ . These have the same dimension as their two-dimensional (string) counterparts. 
Since the action is non-polynomial the quantization process is hampered considerably. 
For this reason we must look for another 
option and supersymmetrize  the Dolan-Tchrakian action at the expense of introducing the $3D$ world volume 
gauge field of dilations $b_\mu$ and relinquishing $S$ supersymmetry and conformal boost invariance as well.

\bigskip
\centerline {\bf 2.3 D-branes, Skyrmions and BPS monopoles} 
\bigskip

In this section we shall discuss the parallels among membranes, Skyrmions , $D$ branes and BPST states. It is not essential for the main results of this work and it may be skipped in the first reading.  

The Dolan-Tchrakian action for the membrane is equivalent to the generalized Skyrme  action 
discussed among others by Manton [11]. We will follow Manton's work closely . 
A Skyrmion may be regarded as a topologically non trivial map from one Riemannian 
manifold to another minimizing a particular energy functional; i.e. classical static field configuration of minimal energy in a nonlinear scalar field theory, the pion field. The standard Skyrmion represents the 
baryon and has a conserved topological charge which precisely prevents a proton from decaying into pions. 
The charge is identified with the conserved baryon number or the degree of the map from 
$R^3 \rightarrow SU(2)$.  

Manton has emphasized that there  is no need for the target manifold to be a Lie Group. Take a map  
$\pi$ from $\Sigma_0 \rightarrow \Sigma_1$. Assume that both base and target space are three dimensional.
The three frame vectors $E^i_m, m=1,2,3 $ on $\Sigma_0$  are mapped by $\pi$ to the vectors 
$E^i_m \partial _i \pi^\alpha =
E^\alpha_m $ on $\Sigma_1$. 
The quadratic terms of the Skyrme model/ Dolan-Tchrakian action ( up to a minus sign) are just
the measure of how the sum of the squared-lenghts of frame vectors changes under the map $\pi$. 
The quartic terms corresponds of how the norm-squared of the area-elements constructed from the 
frame vectors change under the map. The equivalence is established once the corresponding indices are 
properly matched as :

$$ g^{mn} E^i_m \partial _i \pi^\alpha E^j_n \partial _j \pi^\beta g_{\alpha \beta} 
            \leftrightarrow g^{\mu\nu}\partial_{\mu} X^{i}(\sigma)\partial_{\nu} X^{j}(\sigma)\eta_{ij}. 
\eqno (2-14a)$$
similarly the norm of the area elements corresponds to the quartic terms :

$$|E^i_m \partial _i \pi^\alpha \wedge E^j_n \partial _j \pi^\beta|^2 
\leftrightarrow g^{\mu\nu} g^{\rho\tau}\partial_{[\mu} X^{i}(\sigma)\partial_{\rho]} X^{k}(\sigma)
\partial_{[\nu} X^{j}(\sigma)\partial_{\tau]} X^{l}(\sigma)\eta_{ij} \eta_{kl}.\eqno (2-14b)
$$
Eqs-(2-14) have a similar structure to the bosonic terms ( excluding the zero modes ) of the 
lightcone spherical supermembrane moving in a flat target spacetime background : a Yang-Mills theory of the 
area-preserving diffeomorphisms dimensionally reduced to one temporal dimension : a matrix model [2,12].  
The area-squared 
terms are just the same form as the $ [X^I,X^J]^2$ elements  appearing in the light-cone spherical bosonic membrane 
action whereas the length squared terms correspond to the 
ordinary kinetic energy terms. The main difference with eqs-(2-14a, 2-14b) is that $no$ gauge has been fixed, Lorentz covariance is fully explicit and manifest.

Manton has discussed as well how to generalized Skyrme
model to $SU(N)$ for example. The large $N$ limit has been studied by us  
and shown to correspond to a {\bf QCD}  induced  membrane [27] .   
Duff et al [11] had also discussed in the past the behaviour of the actions like (2-7) and (2-13) which violated 
Derrick's theorem. Stable, static, nonsingular, finite energy kink  solutions exist in $3D$ once 
nonstandard actions like the Skyrmion action are built.

Now we turn to the BPS states. As remarked by Manton et al [11] , in recent work,  
there is mounting evidence that there is a close connection between $SU(2)$ BPS monopoles and Skyrmions. 
BPS monopoles are 
solutions of the Bogomolny equation that minimize 
classical energy solutions to the Yang-Mills-Higgs theory.  
Many low energy solutions to Skyrme's  equation look like monopoles with the baryon number 
identified with monopole number. The fields are not the same but the energy density configurations have 
equivalent symmetries 
and approximately the same spatial distribution . 

There is also a deep connection between BPS states and D-branes [9]. $D$-branes are essentially topological defects ( domain walls) where open-strings ends can move. Both in type II and heterotic string 
one can find BPS states in the perturbative string spectrum. 
It is believed that all perturbative string theories are diffferent faces of one underlying theory 
. This is known as string duality [10], where all these theories are believe to be different 
perturbative expansions of one underlying 
theory around different points in the moduli space of string vacua : 
In particular, type II superstring theories have supersymmetric $p$-brane solitonic solutions supported by 
Ramond-Ramond charges that can be reinterpreted as open strings ending on a $p$-brane with Dirichlet 
boundary conditions, the so-called D-brane [9] . D-branes provide a powerful tool to study 
nonperturbative properties of superstring theories and also admit a second class of BPS states that appear 
after compactification of type II and type I strings on a Calabi Yau space. Furthermore, 
there are world volume vector fields in  D-branes actions.

To be more precise, the low energy physics, see reference [13] for an extensive review ,  of $N$ $coincident$ Dirichlet $p$-branes in $10D$ flat space is described in the static gauge ( identifying the world-volume coordinates of the $D$-brane with $p+1$ of the ten dimensional flat space coordinates) by the dimensional reduction to $p+1$ dimensions of a $U(N)~N=1$ SYM in 
$D=10$. $1/2$ of supersymmetries are broken and consequently there are $8$ on-shell bosonic and $8$ on-shell fermionic degrees of freedom. Classically it is a well defined theory but quantum mechanically is anomalous. In particular, the low energy dynamics of $N$ $D0$-branes  
in the gauge where the gauge field $A_0=0$ is given by the dimensionally reduced SYM theory from $10D$ to $0+1$ dimensions. The Lagrangian is  :

$${1\over 2g\sqrt {\alpha '} } Tr [ (\partial_t X^I)(\partial _t X_I) + {1\over (2\pi \alpha')^2} 
\sum [X^I, X^J] ^2 + {1\over 2\pi \alpha'} i\theta^T (\partial_t \theta) -{1\over (2\pi \alpha ')^2} 
\theta^T \Gamma_I [X^I, \theta] ]. \eqno (2-15)$$
Each of the nine adjoint scalar ( from the point of view of the world volume)  matrices $X^I$ is a hermitean $N\times N$ matrix, where $N$ is the number of $0$-branes. The $\theta$ are $16$-component spinors which transform under the $SO(9)$ Clifford algebra given by the $16\times 16$ matrices $\Gamma_I$. 

The $N\rightarrow \infty$ limit of  (2-15) ( see [14] for an update and [15] for a textbook )  can be obtained by replacing the adjoint scalar matrices $X^I$ by $c$-number functions $X^I(t,\sigma^1,\sigma^2)$ and matrix-commutators by Poisson brackets w.r.t two internal coordinates $\sigma^1,\sigma^2$ and the group trace 
by an integral w.r.t the internal coordiantes $\sigma^1,\sigma^2$. In this fashion one obtains a gauge theory of symplectic diffs of a two-dim surface, a membrane. 
The time integral in conjuction with the integration  w.r.t the $\sigma^1,\sigma^2$ variables yields in effect an action similar to the light-cone gauge action ( excluding the zero modes) for a supermembrane of spherical topology moving in a flat target spacetime background. Concentrating solely on the bosonic sector, the $N\rightarrow \infty$ limit of 
(2-15) yields :

$$ lim_{N\rightarrow \infty}  S_{Bosonic} =\int dt \int d^2\sigma ~  {1\over 2g\sqrt {\alpha '} } [ (\partial_t X^I)(\partial _t X_I) + {1\over (2\pi \alpha')^2} 
\sum \{X^I, X^J\} ^2] . \eqno (2-16). $$
We see  that (2-16) is similar to the light-cone gauge spherical membrane action in flat $D=11$. The latter action ( excluding the zero modes) is equivalent to a $10D$ YM action for the $SU(\infty)$ group dimensionally reduced to one temporal dimension. Whereas, (2-16) represents the low energy dynamics of $N=\infty$ coincident $DO$-branes  and was obtained from a dimensional reduction of a $10D$ YM theory to one temporal dimension, after  fixing the gauge $A_0=0$. The $9$ scalars $X^I (t,\sigma^1,\sigma^2)$ are transverse to the world volume of 
the $N=\infty$ coincident $D0$-branes  in a $10D$ flat space and are obtained from  the decomposition of the 
$10D$ YM field $A_M$ into a $p+1$-dim gauge field $A_\mu$ and $10-(p+1)=9-p$ transverse scalars. In the special case that $p=0$ the gauge field $A_\mu=A_0$ and the $9-0=9$ scalars are the $X^I$ transverse coordinates of the $D0$-branes.     
To sum up, the low energy dynamics of an infinite number of coincident $D0$-branes in flat $10D$ space resemble those of a lightcone spherical membrane in flat $11D$ space.  The membrane ground state can be ``seen''  as a condensation of an infinity of $D0$-branes. Since the action given eq-(2-16) has a similar  form to the generalized Skyrmion action of eqs-(2-14a,2-14b) we suggest  that $D0$-branes ought to play an important part in the spectrum of the Skyrmion action in the large $N\rightarrow \infty$ limit.

Solutions to actions of the form (2-16) in $8D$ have been studied by Ivanova-Popov  [16] and Fairlie, Ueno [17] ( octonionic Nahm equations)   
and in particular can be reducced to the ordinary Nahm equations in $4D$ which admit 
BPS monopole solutions. Massive BPS states appear  in theories with $extended $ supersymmetry but even in the case of $N=1$ supersymmetry there is an analog of BPS sates, namely the massless 
states [18]. A future proyect is to see what connection there may be, if any,  among 
these massless sates, the analogs of BPS states, with the infinite number of coincident $D0$-branes and the spectrum of the generalized Skyrmion action.    
Hints that a relation exists are based on the fact that :
(i) we have a world-volume vector field $b_\mu$ in our Weyl invariant spinning membrane action 
in contrast with the 
$p+1$-dim $U(N)$ gauge field $A_\mu$ living on the world-volume of a $Dp$-brane.   
(ii) Weyl conformal invariance (iii) $Q$  supersymmetry only ( 1/2 supersymmetry, as it occurs in BPS states) and no 
$S$ supersymmetry.     
(iv) The Skyrmion action (2-14a,2-14b) subsumes the action for the large $N\rightarrow \infty$ coincident $D0$-branes given by eq-(2-16).  
  
For work related to the world volume supersymmetry of Born-Infel actions see [28] and for Non Abelian Born-Infeld Skyrmions see [29]. On other brane actions and superembeddings see [26].

\bigskip

\centerline{\bf { III}. The ${\cal Q} $-spinning Weyl Invariant Membrane}
\bigskip
\centerline{\bf 3.1 Supersymmetrization of the Kinetic Terms}
\bigskip

    In this section we will present an action for the $3D$ Kinetic matter
superconformal multiplet where supersymmetry is linearly realized and without $R$ terms.
In particular, we will show why the action is invariant under a $modified$  ${\cal Q} $-transformation , 
${\tilde \delta}_Q = \delta_Q+ \delta'$, which includes a compensating $S$ and $K$-transformation which cancel the anomalous/spurious contributions due to the explicit breakdown of $S$ and $K$-invariance of the action. 
The latter action is invariant under  $ P, D, M^{ab}$ transformations :  translations, dilations, Lorentz.  
But is $not$ invariant under conformal boost  $K$ and $S$-supersymmetry. We will show that the gauge algebra closes on the action : $[{\tilde \delta}_Q, {\tilde \delta} _Q ] {\cal S} = 
{\bar \epsilon^2} \gamma^\mu \epsilon^1 {\cal D}^c_\mu {\cal S} =0$ ; 
where ${\cal D}^c_\mu$ is the $modified$ supercovariant derivative operator obtained after a 
compensating transformation $\delta'$ is added to the standard $Q$-transformation to cancel the 
anomalous $S$ and $K$-transformations of the supermultiplets used to construct the action.  
It will be explained in full detail what is the difference between the ordinary $S,K$-transformations and the anomalous ones ( acting on the action ). 

What one has is then a $Q+K+S$ sum rule for the 
$modified$  ${\cal Q} $ transformation : ${\tilde \delta}_Q = \delta_Q+ \delta'$. 
The closure of two succesive modified ${\tilde \delta_Q} $ transformations occurs despite the fact the action explicitly breaks conformal boosts $K$ and $S$-supersymmetry, 
$\delta_K {\cal S } \not=0;~\delta_S {\cal S} \not=0$. However, the combination of a 
$\delta_K {\cal S }$ variation with a $ \delta_S {\cal S}$ variation, encoded inside the $\delta' {\cal S} $ variation, 
$cancels$ the anomalous variation   
$\delta_Q {\cal S}\not=0 $, as we intend to prove in this section. 
 
Unfortunately, the authors in [27] left out completely the crucial $ compensating $ role ( to the explicit $S$-breaking terms of the action ) that the breakdown of conformal boost symmetry had in the construction of the following actions. 
The actions below explicitly break conformal boosts due to the explicit appearance of the gauge field of dilations 
$b_\mu$ which is $not$ inert under conformal boosts. The scalar Weyl multiplet is inert under $K$-transformations and so are the $e^m_\mu, \psi_\mu$. Despite the explicit presence of $b_\mu$ terms ( after unravelling the covariant derivatives ) we will also show that the action is Weyl invariant.

The modified ${\cal Q}$-supersymmetry is reminiscent of the $Q+S$ sum rule of 
Poincare Supergravity actions. They are $not$ separately invariant under $Q$ nor $S$-supersymmetry. However, they are invariant under the combination : $\delta_Q + \delta_S (\epsilon_S = -F \epsilon_Q )$; i.e, the 
$\epsilon_S = -F\epsilon_Q$ is a field-dependent fermionic parameter.  Despite that the $Q$ and 
$S$-variations are independent due to the field-dependent $\epsilon_S$, they become entangled and `` add `` up to $zero$. Something analogous happens here. Roughly speaking one has here a $Q+K+S$-sum rule with the main 
difference that one is not fixing by hand any symmetry : the action itself is the one which explicitly breaks those symmetries. It is important to emphasize that the action is invariant under the modified 
${\cal Q}$-transformations, while it is $not$ invariant under the ordinary $Q$-supersymmetry.

Also we will supersymmetrize the quartic-derivative terms of (2-7). This is
attained by using directly an explicit  superconformally invariant action for
the kinetic terms. The quartic terms do not admit a superconformally invariant
extension unless one includes a suitable coupling `` constant `` as we shall see shortly. The key issue lies in the fact that if we
wish to satisfy the three requirements: 

1). A spinning membrane action which is polynomial in the fields.

2).  Absence of $R$ terms and kinetic gravitino terms.

3). Linearly realized supersymmetry in the space of fields after the
elimination of the auxiliary fields, before and after one sets the Fermi fields
to zero. 

Then, one must relinquish $S$ supersymmetry $and$ conformal boosts symmetry altogether and concentrate solely on the
( modified ) ${\cal Q} $ supersymmetry, translational, dilations and Lorentz symmetries associated with a subalgebra of the full the superconformal algebra in D=3. We shall
begin with some definitions of simple-conformal SG in D=3 [4]:

The scalar and kinetic multiplet of simple conformal SG in $3D$  are respectively:

$$\Sigma_c =(A,\chi,F).~~T_c(\Sigma_c)=(F,\gamma^\mu D_\mu^c \chi, \triangle  A)\eqno(3-1){ }$$

We have the following quantities:

$$ D^c_\mu A =\partial_\mu A -{1\over 2} {\bar \psi}_\mu\chi -\omega (A)  b_\mu A.                                              \eqno(3-2a)$$

$$  D^c_\mu \chi =(D_\mu -(\omega (A)  +{1\over 2})b_\mu )\chi -{1\over 2}\gamma^\mu D_\mu^c A\psi_\mu 
-{1\over 2}F\psi_\mu -\omega(A) A\phi_\mu.
\eqno(3-3a)$$
where the Weyl weight of the first member of the conformal supermultiplet is $\omega (A)$.

The  superconformally covariant derivative of fields $A, \chi$ are obtained from the commutator of two infinitesimal $Q$-transformations, with fermionic parameters, 
$\epsilon^1_Q, \epsilon^2_Q $ respectively,  acting on $A,  \chi$ . Such commutator is naturally given by  is a combination of the superconformal symmetries:

$${\bar \epsilon}^1_Q \gamma^\mu \epsilon^2_Q  D^c_\mu A = \sum_A c_A \delta_A (\epsilon_A) A \eqno ( 3-2b) $$ 
$${\bar \epsilon}^1_Q \gamma^\mu \epsilon^2_Q  D^c_\mu \chi   = \sum_A c_A \delta_A (\epsilon_A) \chi \eqno (3-3b)   $$ 
where the parameters $\epsilon_A$ are associated with  $all$  the elements of the superconformal algebra. Since the $A, \chi$ are $inert$ under conformal boosts the relevant generators in the r.h.s of eqs-(3-2b, 3-3b) are the infinitesimal translations, Lorentz, dilations, $Q$ and $S$ supersymmetries with the following corresponding $field-dependent$  parameters :

$$\epsilon^a _P =  {\bar \epsilon}^1_Q \gamma^\mu \epsilon^2_Q e^a_\mu. ~~~
\epsilon^{ab}_{Lorentz}  =  {\bar \epsilon}^1_Q \gamma^\mu \epsilon^2_Q \omega^{ab}_\mu. ~~~
\epsilon_D =  {\bar \epsilon}^1_Q \gamma^\mu \epsilon^2_Q b_\mu. \eqno (3-4a) $$   

$$ \epsilon^\alpha_Q =  {\bar \epsilon}^1_Q \gamma^\mu \epsilon^2_Q \psi_\mu. ~~~
\epsilon^\alpha_S =  {\bar \epsilon}^1_Q \gamma^\mu \epsilon^2_Q \phi_\mu. ~~~ 
\xi^m_K =  {\bar \epsilon}^1_Q \gamma^\mu \epsilon^2_Q f^m_\mu .          \eqno (3-4b)$$
where the last two parameters represent the field-dependent $S$ and $K$-transformations. $\phi_\mu$ is the gravitino field-strength and $f^m_\mu$ is comprised of the curvature scalar $R$ and kinetic gravitino terms.

Notice how important is to have the right index structure in eqs-(3-4). $Loosely$ speaking, one can  say that the superconformal derivative    $ D^c_\mu \chi$  is obtained as a result of an $S$ supersymmetry transformation with an $\epsilon_S = \phi_\mu$ ( the gravitino field strength ) . However, rigorously speaking $\phi_\mu$ by itself does not qualify for the correct $\epsilon_S$.  
Because $\phi_\mu$ has both vectorial and spinorial indices it needs to be coupled to the correct 
$  {\bar \epsilon}^1_Q \gamma^\mu \epsilon^2_Q$ in order to qualify for an $\epsilon_S$. $ D^c_\mu \chi$ is obtained as a combination of all the symmetries with the suitable $field-dependent$ parameters given by eqs-(3-4).

The source of this argument in expressing the superconformal covariant derivative acting on the 
fields as a commutator of two infinitesimal $Q$-supersymmetries is related to the expression of 
the Lie-superalgebra valued connection  : ${\cal A}_\mu = {\cal A}_\mu ^A X_A$ and the Lie-algebra of derivations. The connection  
admits an expansion into the three independent gauge fields : 
$e^a_\mu P_a  $ ( gauges translations ); ${\bar \psi} _\mu Q $ ( gauges $Q$-supersymmetry ); 
the dilatation field $b_\mu D $ 
. Whereas the remaining fields appearing in the expansion : spin connection ( gauges Lorentz ) $\omega^{ab}_\mu M_{ab}$ ; 
the $f^a_\mu $, which contains explicitly the scalar curvature $R$ and the kinetic terms for the gravitino couples to the conformal boosts generator $K_a$; 
and the $\phi_\mu$ ( gravitino field strength )   
couples to the $S$-supersymmetry generator . The latter three are not truly independent fields but are solved in terms of the fundamental ones $e^a_\mu, \Psi_\mu, b_\mu$ , after imposing  the standard three constraints  on the group curvatures .

At first sight one may be inclined to say that it is imposible to relinquish $S$ supersymmetry without a 
gauge fixing procedure. However, we must $not$  forget that the conformal boost symmetry is also explicitly broken by our action. This is precisely why the $b_\mu$ fields do $not$ decouple from the action, as it occurs in the usual full superconformaly invariant actions . We will show, contrary to what was argued by some authors that the gauge algebra of two $modified$ ${\cal Q} $-transformations, $closes$ on the action :

$$   [ {\tilde \delta}_Q (\epsilon^2),~    {\tilde \delta}_Q (\epsilon^1) ]= {\bar \epsilon}^1_Q \gamma^\mu \epsilon^2_Q  {\cal D}^c_\mu {\cal S} = \sum_A c_A \delta_A (\epsilon_A) ~{\cal S}= 0  \eqno ( 3-5) $$ 
Where    ${\cal D}^c_\mu$ is the $modified$ derivative obtained by cancelling  the anomalous $S,K$ transformations , by adding suitable compensating transformations $\delta' $ to the ordinary $Q$ transformation; ${\tilde\delta_Q}   =   \delta_Q + \delta' $.

We shall prove in {\bf 3.2} that the gauge algebra closes on the ${\cal Q}$-invariant Weyl spinning membrane actions, 
despite the fact the latter breaks both $S, K$ symmetries and the standard $Q$ supersymmetry explicitly  : 

$$   \delta (\epsilon_Q) {\cal S} \not= 0.~~~ \delta (\epsilon_S) {\cal S} \not= 0.~~~\delta (\epsilon_K) {\cal S} \not=0. \eqno (3-5a) $$
nevertheless their combination in the $Q+K+S$ sum rule nevertheless `` adds `` up to zero :  

$$   {\tilde \delta}_Q {\cal S} = c_Q\delta (\epsilon_Q) {\cal S} + c_S\delta (\epsilon_S) {\cal S} + c_K \delta (\epsilon_K) {\cal S} =0 .\eqno (3-5b) $$ 
which is a condition required from the closure of the gauge algebra. Hence,  
if, and only if, the action is 
${\cal Q} $-invariant, and invariant under all the other symmetries, $except$ the $S$ and $K$-symmetries ( conformal boosts ) , the condition ${\tilde\delta}_Q {\cal S} = 0 $ automatically implies that the commutator of two successive ${\tilde Q} $'s yields also  zero :

$$ If~~~{\tilde\delta}_Q {\cal S} = 0 \Rightarrow  
[{\tilde \delta} ^{\epsilon_1}_Q, {\tilde\delta}^{\epsilon_2}_Q ] \int d^3x~  [ e {\cal L} ] =0 \Rightarrow   
{\bar \epsilon}^1_Q \gamma^\mu \epsilon^2_Q {\cal D}^c_\mu \int d^3x~ [ e {\cal L} ] = 0 \Rightarrow $$
$$\sum_A c_A \delta^{ \epsilon_A }_A 
\int d^3x~ [e {\cal L}]=  ( \sum_{unbroken} c_A \delta^{ \epsilon_A }_A [{\cal S}] )~  +    
[ c_Q \delta_Q^{\epsilon_Q} + c_S \delta^{ \epsilon_S }_S +  c_K\delta^{ \epsilon_K }_K ] [{\cal S}]  = 0. \eqno (3-6)$$
Since the unbroken symmetries yield a zero variation, from the last three terms of eq-(3-6) we can infer that the $combination$  of the anomalous $Q, S,K$ transformations acting on the action `` adds `` up to zero as required. 
It was unfortunate, that some authors  $left$ out completely the possible `` compensating'' role of the conformal boosts symmetry, relative to the $S$-supersymmetry, in the closure of gauge algebra on the action in eq-(3-6). 
The closure of the gauge algebra on the action entails that the action is translational, dilational, Lorentz and ${\cal Q}$-invariant but breaks $Q, S,K$-invariance in such a way that the combined  effect of the latter two breaking symmetries cancels the anomalous ordinary $Q$ variations of the action , as shown in eqs-(3-5, 3-6).

The apparent objection one could have to this argument is the following : In general, 
the variations $\delta_Q {\cal S}; \delta_K {\cal S} $ and $\delta_S {\cal S} $ are independent. How is it possible that their 
combination cancels out exactly giving zero ? The answer is given precisely by the fact that the 
$S$ and $K$-transformations appearing in the r.h.s of (3-6) ( the modified covariant derivative) are given in terms of $field-dependent$ parameters : $\epsilon_S, \epsilon_K$ 
eqs-(3-4a, 3-4b ).  As a result, there is going to be a non-trivial $entanglement$  of 
the variations  $\delta_Q {\cal S}; \delta_K {\cal S} $ and $\delta_S {\cal S} $ ; i.e one does not longer have a $linear$ combination of formally independent variations.

To illustrate how this is possible we present the most relevant known example : Poincare supergravity. 
 For example, the Poincare Supergravity action is not $separately$ invariant under $Q$ nor $S$ supersymmetries. However, it is invariant under a suitable linear combination of the two , the so-called sum rule $Q+S$. 
Why ? because  
the $S$-fermionic parameter is $field-dependent$ : $ \epsilon_S = -F \epsilon_Q. $, therefore the  $Q$ and $S$ symmetries are $entangled$ and the linear combination of two formally independent transformations 
can yield  a net value of zero.  
The modified ${\tilde  Q}$-transformation is essentialy a sort of $Q+K+S$-sum rule. If the parameters had not been 
field-independent, no $entanglement$ relation between the $Q; S$ and $K$ transformations would have been possible, and the only possible solution would have been : $c_Q = c_S= c_K = 0$; i.e a truly linear-independence between $S$ and $K$ transformations.

How about the closure on the gauge fields ? $(e^a_\mu, b_\mu, \psi_\mu, f^m_\mu, \omega^{ab}_\mu, \phi_\mu)$. Since these fields transform properly under $S, K$-transformations there is not need to subject them to the modified ${\cal Q}$-transformations.  The algebra closes on the gauge fields.

After this important detour  
following the same arguments that a supercovariant derivative is built from a combination of all the symmetries we arrive at the superconformal D\ Alambertian : 

$$\triangle  A =D^c_a D^{ca} A=e^{-1} \partial_\nu (eg^{\mu\nu}D^c_\mu A)+
{1\over 2}{\bar \phi}_\mu \gamma^\mu \chi -[\omega (A) - 1]  b^\mu D^c_\mu A +$$
$$2\omega(A)  A f^a_\mu e^\mu_a -{1\over 2}{\bar \psi}^\mu D^c_\mu \chi 
-{1\over 2}{\bar \psi}^\mu \gamma^\nu \psi_\nu D^c_\mu A. \eqno(3-7){ }$$
where : $\omega (A) $ is the Weyl weight of the scalar $A$. Meaning, that under Weyl transformations it obeys
$\delta_{Weyl } A = \omega (A) \lambda A. $ We corrected the 
$[\omega (A) - 1]  b^\mu D^c_\mu A$ term in (3-7). A factor of $- b^\mu D^c_\mu A $ was missing in [4]. 

Notice how the term containing the curvature scalar and the kinetic gravitino term, $f^a_\mu$ is the one that couples to a $non-trivial $ conformal boost symmetry of the $b_\mu$ field apperaing in the definition of 
$\xi^\mu D_\mu^c A $ eq-(3-2a). Therefore, a $K$-variation of the  $D^c_\mu A$, required in the D'Alambertian, yields the required term containg the curvature  

$$\delta_K ( \epsilon_K) b_\mu = - 2 \epsilon^a_K e_{a\mu} =  
-2 [ {\bar \epsilon}^1_Q \gamma^\nu \epsilon^2_Q f^a_\nu  ]  e_{a\mu}  . \eqno (3-8) $$  
this non-trivial boost transformation of the $b_\mu$ is the one responsible for the presence of curvature terms in the superconformal D'Alambertian. Such curvature terms did not appear in the definition of the 
supercovariant derivative of $A, \chi$ because these fields are $inert$  under conformal boosts.

The generalized spin connection is :

$$\omega^{mn}_\mu =-\omega^{mn}_\mu (e)-\kappa^{mn}_\mu (\psi)+e^n_\mu b^m -e^m_\mu b^n.
\eqno(3-9-a)$$

$$\kappa^{mn}_\mu ={1\over 4} ({\bar \psi}_\mu \gamma^m \psi^n -{\bar \psi}_\mu \gamma^n \psi^m 
+ {\bar \psi}^m \gamma_\mu \psi^n). 
\eqno(3-9-b)$$
The gravitino field strength is : 

$$\phi_\mu ={1\over 4} \gamma^\lambda \gamma^\sigma S_{\sigma \lambda} =
{1\over 4} \sigma ^{\lambda\sigma}  \gamma_\mu S_{\sigma \lambda}. 
\eqno(3-9-c) $$

where :  
$$S_{\mu\nu} =(D_\nu +{1\over 2} b_\nu )\psi_\mu - \mu \leftrightarrow \nu. 
\eqno(3-9-d) $$

$$e^{a\mu} f_{a\mu} =-{1\over 8} R(e,\omega) -{1\over 4}{\bar \psi} _\mu \sigma^{\mu\nu}\phi_\nu. 
\eqno(3-9-e)$$

The transformation laws under Weyl scalings of the $canonical$  supermultiplet and $e^m_\mu$ are respectively :

$$ \delta e^m_\mu = - \lambda e^m_\mu.~~~\delta A={1\over 2}\lambda A.~~~\delta \chi=\lambda \chi.~~~
\delta F={3\over 2}\lambda F
\eqno(3-10-a )$$
meaning that the Weyl weights,  in units of the fiduciary Weyl parameter $\lambda$,  of the fields $e^m_\mu; A, \chi, F $ are respectively $-1; (1/2); 1; (3/2)$.

For example, a supermultiplet whose Weyl weight ( its first component)   
is $\omega (A)=\omega$ transforms as :

$$\delta A=( \omega )\lambda A.~~~\delta \chi= (\omega+ {1\over 2} )\lambda \chi.~~~
\delta F=(\omega + 1)\lambda F. \eqno (3-10-b)$$
  
Having specified what one means by Weyl weights, hopefully will clarify another source of misunderstanding 
[27 ].  The notation used by Uematsu was a little bit misleading in the sense that he refers to the 
Weyl weights $\lambda$ , indistinguishably, as the fiduciary Weyl parameter $\Lambda$, that is 
written also as $\lambda$, in the Weyl scalings associated with the $canonical$  multiplet appearing in eq-(3-10-a),  
as well as ordinary $numbers$ in the definitions of the supermultiplets and derivatives, etc......  

The $Q$ and $S$-supersymmetry are
respectively:

$$\delta^c_Q A ={\bar \epsilon} \chi.~~~\delta^c_Q \chi =F\epsilon+\gamma^\mu D^c_\mu A\epsilon. ~~~
\delta^c_Q F={\bar \epsilon}\gamma^\mu D^c_\mu \chi.
\eqno(3-11)$$

$$\delta^c_Q e^m_\mu ={\bar \epsilon}\gamma^m \psi_\mu.~~~
\delta^c_Q \psi_\mu  =2( D_\mu +{1\over 2}b_\mu )\epsilon. ~~~
\delta^c_Q b_\mu = \phi_\mu.
\eqno(3-12 )$$
The $S$-supersymmetry transormations are :

$$\delta^c_S e^m_\mu =0.~~~ \delta^c_S \psi_\mu =-\gamma_\mu \epsilon_s.~~~ 
\delta^c_S b_\mu =-{1\over 2}\psi_\mu \epsilon_s.  \eqno (3-13-a) $$       

$$\delta^c_S \omega^{mn}_\mu =- {\bar \epsilon}_s \sigma^{mn} \psi_\mu.$$

$$\delta^c_S A =0. ~~\delta^c_S \chi =\omega(A)  A \epsilon_s.~~~
\delta^c_S F = ({1\over 2}-\omega (A) ) {\bar \chi} \epsilon_s. \eqno (3-13-b)$$

The canonical scalar supermultiplet is $inert$ under conformal boots $K$  and so are $e^m_\mu, \psi_\mu$. 
The $b_\mu$ is $not$  inert : 

$$\delta_K (\xi^m_K ) b_\mu = - 2 \xi^m_K e_{m \mu}.~~~   \delta_K (\xi^m_K ) \omega^{mn}_\mu = 
    2 (  \xi^m_K e^n_\mu -    \xi^n_K e^m_\mu).         \eqno (3-13c )  $$

Notice that the scalar multiplet (3-11) transforms properly under
$Q$ transformations for any value of the conformal weight but $not$  under $S$  supersymmetry 
transformations unless one assigns the canonical weight  
$\omega (A)  ={1\over 2}$ to the $\Sigma_C$ supermultiplet ( to its first member, $A$ )   
so that the first component of its associated kinetic multiplet : $T_c (\Sigma_c)$ ,  
${\cal A}^T \equiv  F$ has the correct $S$-transformation law :

$$ \delta_S {\cal A}^T = \delta_S F = ( {1\over 2} - \omega (A) ) {\bar \chi} \epsilon_S =  0          $$
This means that  the associated  kinetic  $T_c (\Sigma_c)$
multiplet to the canonical multiplet has a Weyl weight equal to the weight of $F= 1+{1\over 2}$.   

For example, in two-dimensions the kinetic multiplet $ T_C ( \Sigma_C \otimes \Sigma_C)$ does transform properly under $S$ transformations because the Weyl weight of the canonical supermultiplet is $zero$. In $D = 2$ the $S$ transformation $differ$ from the ones in $D = 3 $ : 

$$\delta^c_S A =0. ~~\delta^c_S \chi =\omega(A)  A \epsilon_s.~~~
\delta^c_S F =  -\omega (A)  {\bar \chi} \epsilon_s. ~~~D=2. \eqno (3-13-d)$$

Since the Weyl weight of the canonical supermultiplet in $ D = 2$ is $zero$ this means that it is $inert$  
under $S$-transformations and hence  $ T_C ( \Sigma_C \otimes \Sigma_C)$ transforms properly. 
However, this is $not$ the case for $D = 3$. 
Such kinetic multiplet $breaks$ explicitly $S$ and conformal boosts. 
The first component of the  
$T_C ( \Sigma_C \otimes \Sigma_C)$ is $ A_1 F_2 + A_2 F_1 - {\bar \chi}_1 \chi_2 $ and such 
composite-component $does~ not$ transfom properly under $S$ supersymetry : 

$$  \delta_S~( A_1 F_2 + A_2 F_1 - {\bar \chi}_1 \chi_2  ) = - \delta_S~ ( {\bar \chi}_1 \chi_2)  =
- {1\over 2} A_2 {\bar \chi}_1 \epsilon_S   - {1\over 2} A_1 {\bar \chi}_2 \epsilon_S \not=0. \eqno (3-14)$$

In $D=2$ the $S$-transformation of this composite-component is $zero$  as it ought to be, due to the fact that the Weyl weight of the canonical multiplet is zero so the $\chi$ fields are inert under $S$-transformations.

For this reason, a  superconformally invariant action for the kinetic terms in $D=3$ requires to include $only$  the    
following kinetic multiplet that is fully superconformaly invariant : 
$\Sigma_C\otimes T_C(\Sigma_C)$. The superconformal invariant action is : 

$${L=e[{\hat F}+{1\over2}\bar{\psi_\mu}\gamma^\mu {\hat \chi}
+{1\over2}{\hat A}\bar{\psi_\mu}\sigma^{\mu\nu}\psi_\nu]}. ~~~D=3. \eqno(3-15)$$  
where one inserts the multiplet 
$\Sigma_C\otimes T_C(\Sigma_C)=( {\hat A}, {\hat \chi} ,{\hat F})$ into (3-15). The explicit components of 
the latter multiplet are :

$$ {\hat A}=AF.~~~{\hat \chi}=A\gamma^\mu D^c_\mu \chi +F\chi.~~~
{\hat F} =A\triangle A- {\bar \chi} \gamma^\mu D^c_\mu \chi +F^2. ~~~~D = 3. \eqno (3-16)$$

Had one $added$  the other kinetic multiplet $T_C ( \Sigma_C \otimes \Sigma_C)$ to eqs-(3-15, 3-16) , 
one would have broken explictly the $S$ and conformal boost symmetries while maintaing the other symmetries intact;  i.e the integration by parts that allows to rewrite : 
$ A_i (\partial^\mu \partial_\mu A_j) \Rightarrow (\partial_\mu A_i)(\partial_\mu A_j) $ breaks explicitly the $S$ and $K$-symmetries. The $T$ operation in $T_C ( \Sigma_C \otimes \Sigma_C)$ is roughly nothing but an integration by parts. Such $T$-operation does $not$ `` commute ``  with the $\delta_S, \delta_K$ operations  in $D = 3$.

The reason why one  must have the correct Weyl weight $\omega (A)   = {1\over 2}$ for $\Sigma_C$ so that the ${\hat F} $ 
component 
appearing in (3-15 ) has dimension three, is because otherwise we would not
even have $Q$-invariance in the action, despite the fact that the kinetic multiplet transforms
properly under $Q$ -transformations irrespectively of the values of $\omega$.
On physical grounds we see that the notion of canonical dimension is
intrinsically tied up with the conformal invariant aspect of the kinetic terms
in the action. We have a conformally invariant kinetic term if, and only if, the
fields have the right (canonical) dimensions to yield terms of dimension three in
the Lagrangian.

We might ask ourselves how did Lindstrom \& Rocek manage to
construct a Weyl invariant spinning membrane when their fields had a
non-canonical dimension? The answer to this question lies on the nonpolynomial
character of their action. Formally one has an infinite series expansion where the whole sum of 
explicitly $Q$ and $S$ supersymmetry breaking terms  is effectively invariant under the $''Q+S''$ 
sum rule.  
An example of a multiplet that transforms properly under the $''Q+S''$ sum rule but $not$ separately 
under $Q$ nor $S$ supersymmetry is the following Poncare kinetic multiplet :

$$T_p(\Sigma_p)=(F; D^c\chi (\omega ={1\over2}); \triangle A -{3\over 4}FS
).\eqno (3-17)$$
This multiplet is almost `' identical ``  as the kinetic superconformal multiplet (3-1) 
except that the last component is different due to the presence of the $-{3\over 4} FS$ term. 
We emphasize at this point, that the multiplet of 
eq-(3-17) $is~ not$ the multiplet 
we are going to $use$ in the construction of our actions.  
We will use a $different$ multiplet.

The task now is to see how do we write a suitable action for the kinetic
terms without $R$ terms ( which appear in the definition of the D'Alambertian)
The suitable action is obtained by using the following combinations of supermultiplets 
as follows:

The   combination $\Sigma^i_C \otimes T_C(\Sigma^j_C ) + T_C(\Sigma^i_C
)\otimes \Sigma^j_C -T_C( \Sigma^i_C \otimes \Sigma^j_C)$ 
happens to be the correct one to
dispense of the $R$ and kinetic gravitino terms. However this combination $breaks$ explicitly the $S$ and conformal boosts as we intend to show next.

The explicit components of the suitable combination of multiplets which removes the $R$ and gravitino kinetic terms and breaks $K$-symmetry  and $S$-supersymmetry ,    $\Sigma^i_C \otimes T_C(\Sigma^j_C ) + T_C(\Sigma^i_C
)\otimes \Sigma^j_C -T_C( \Sigma^i_C \otimes \Sigma^j_C)$  are  :                                   :

$$A_{ij}= \bar{\chi_i} \chi_j.\eqno(3-18a)$$                                  

$$\chi_{ij} =F_i\chi_j +F_j\chi_i +A_i\gamma^\mu D^c_\mu (\omega ={1\over 2})\chi_j  
+A_j\gamma^\mu D^c_\mu (\omega ={1\over 2})\chi_i-$$
$$\gamma^\mu D^c_\mu (\omega =1)[A_i\chi_j + A_j\chi_i].        
\eqno(3-18b)$$            

$$F_{ij}= A_i \triangle (\omega ={1\over 2})  A_j +A_j \triangle (\omega ={1\over 2})A_i +2F_iF_j 
-{\bar \chi}_i \gamma^\mu D^c_\mu ( \omega ={1\over 2})\chi_j -$$      
$$  {\bar \chi}_j \gamma^\mu D^c_\mu ( \omega ={1\over 2})\chi_i - \triangle ( \omega =1)[A_iA_j]. 
\eqno (3-18c)$$       

The $F_{ij}$ terms contain the standard kinetic terms : 

$$-2g^{\mu\nu}\partial_\mu A_i\partial_\nu A_j -\bar{\chi_i}\gamma^\mu D_\mu\chi_j
+2F_iF_j+......\eqno (3-18d)$$ 
and $no$ curvature terms by construction. The explicit expression for the 
 $F_{ij}$ ( we added the contribution induced by the factor of $\omega(A) - 1 $ in (3-7) and corrected a minus sign in one term which was brought to our attention by [27] )  is :

$$F_{ij}=-2g^{\mu\nu}\partial_\mu A_i\partial_\nu A_j -
(\bar{\chi_i}\gamma^\mu D^c_\mu (\omega ={1\over 2})\chi_j +i\leftrightarrow j)
+2F_iF_j - {1\over 2} b^\mu b_\mu A_i A_j +$$
$$b_\mu A_i A_j {\bar \psi}^\mu \gamma^\nu \psi_\nu - {1\over 2} b^\mu \partial_\mu ( A_i A_j) + 
{\bar \psi}^\mu (\chi_j \partial_\mu A_i+i\leftrightarrow j) -{1\over 2}
{\bar \psi}^\mu \gamma^\nu \psi_\nu \partial_\mu (A_i A_j)+$$
$${1\over 4}
{\bar \psi}^\mu \gamma^\nu \psi_\nu 
(A_i{\bar \psi}_\mu \chi_j +i\leftrightarrow j). 
\eqno (3-19)$$

The remaining components are :
$$\chi_{ij}={1\over 2}\gamma^\mu b_\mu A_i \chi_j 
-\gamma^\mu\partial_\mu A_i  \chi_j  
+F_i\chi_j + i\leftrightarrow j. ~~~A_{ij}= {\bar \chi}_i \chi_j.\eqno (3-20)$$ 

As promised earlier, we can see that there is no gravitino kinetic term , nor curvature scalar terms, as expected. Notice the explicit presence of the $b_\mu$ terms, signaling an explicit  breaking of conformal boosts.  The ordinary canonical scalar multiplet  was inert under conformal boosts. However, this does $not$ mean that 
the $T(\Sigma\otimes\Sigma)$ is also inert under conformal boosts. 
Therefore, eliminating the $R$ and gravitino kinetic terms is $not$ compatible with
$S$-supersymmetry not $K$-symmetry. We are forced, then, to relinquish $both$ the $S$-supersymmetry and conformal boost symmetry in the explicit form of the action, without fixing them by hand, and to 
implement explicitly a modified ${\cal Q}$ and $ P, D, M^{ab}$ in the action only; 
i.e a supersymmetric extension of the `` homothecy ``  transformations. 
   
Because the component $T(\Sigma\otimes\Sigma)$  does not have the correct $S; K $-transformations laws, 
the components of the latter supermultiplet mutiplet $ (A_{ij}; ~\chi_{ij}; ~ F_{ij} ) $  in eqs-(3-18) $do ~ not$ longer  transform properly under ( ordinary ) $Q$ transformations because the supercovariant derivatives acting on the $A_{ij}, \chi_{ij}$  acquire anomalous/spurious terms due to the explicit $S$ and $K$-breaking pieces. 
The supercovariant derivatives are  built in terms of combinations of $all$ the generators of the full algebra. 
The broken and unbroken symmetries. We shall modify the $Q$-transformation law to compensate for 
the breakdown of $S$-supersymmetry and conformal boost symmetry.

The supersymmetric quadratic derivative terms ( kinetic  action )  is obtained by plugging in directly the components 
$A_{ij},\chi_{ij}, F_{ij}$ obtained in eqs-(3-18, 3-19, 3-20 )  into  the following expression while contracting the spacetime indices with $\eta_{ij}$  :

$${\cal L}_2=e\eta^{ij} [F_{ij} +{1\over 2}{\bar \psi}_\mu \gamma^\mu \chi_{ij} +
{1\over 2}  A_{ij}{\bar \psi}_\mu \sigma^{\mu\nu} \psi_\nu ]. \eqno (3-21)$$
 
Using the modified derivatives, the action will be ${\cal Q} $-invariant under the following modified ${\cal Q} $-transformations :

$$  {\tilde \delta} ^c_Q (A_{ij}) = \delta^c_Q (A_{ij})  = {\bar \epsilon} \chi_{ij}. ~~~
{\tilde \delta} ^c_Q (\chi_{ij}) = F_{ij} \epsilon + \gamma^\mu 
{\cal D}^c_\mu (A_{ij})\epsilon. ~~~ {\tilde \delta} ^c_Q F_{ij} = {\bar \epsilon} \gamma^\mu {\cal D}^c_\mu (\chi_{ij}). \eqno (3-22) $$
where now is essential to use the modified supercovariant derivatives ${\cal D}^c_\mu$ 
associated with the modified ${\cal Q}$-transformations 
laws in order to cancel the anomalous contributions stemming from the $S$ and $K$-breaking pieces, since the 
latter anomalous contributions to the $Q$-variations are contained in the 
$ordinary$ derivatives $D^c_\mu (A_{ij}); ~  D^c_\mu (\chi_{ij})$ .

The modified covariant derivatives, appearing in the modified ${\cal Q}$-transformation laws (3-22) of the supermultiplet (3-18, 3-19, 3-20 ) of Weyl weight $\omega = 2$  used to construct the action (3-21),  have now the appropriate form  :

$${\cal D} ^c_\mu (A_{ij}) = \partial_\mu A_{ij} - {1\over 2} {\bar \psi_\mu} \chi_{ij} - (2) b_\mu A_{ij} . 
\eqno (3-23) $$

$${\cal D} ^c_\mu \chi_{ij}  = ( D_\mu - (2+ {1\over 2} ) b_\mu ) \chi_{ij} - 
{1\over 2} \gamma^\nu {\cal D}^c_\nu (A_{ij}) \psi_\mu - {1\over 2} F_{ij} \psi_\mu - 
(2)  A_{ij} \phi_\mu . \eqno (3-24 )$$

To sum up, after using a modified ${\cal Q}$-transformations (3-23, 3-24 )  to cancel out the anomalous $S,K$-transformations of the supermultiplet given by eqs- (3-18, 3-19, 3-20), used to construct the action (3-21), 
one will obtain an action invariant under the ${\cal Q}$-transformations , given explicitly by eq-(3-23), where one retains in the $modified$ covariant derivative ${\cal D}^c_\mu$ only the non-anomalous pieces and $discards$ the anomalous ones. The modified ${\tilde \delta_Q}$ transformations (3-23)   achieves  precisely that.

\bigskip

\centerline{\bf  3.2 Closure of the Algebra and Additional Remarks } 
\bigskip

This section may be skipped at the first reading. It contains explicit details about the closure of the modified 
${\cal Q}$-susy and additional remarks.

The anomalous transformation laws of the supermultiplet appearing in eqs-(3-18, 3-19, 3-20),  
under the ordinary $Q$-transformations,  will yield 
anomalous/spurious  contributions to the  variations of the Lagrangian density due to the anomalous/spurious terms appearing in the standard supercovariant derivatives :

$$\gamma^\mu D^c_\mu (A_{ij}) = \gamma^\mu [\partial_\mu A_{ij} - {1\over 2} {\bar \psi_\mu} \chi_{ij} - (2) b_\mu A_{ij}] + 
[ c_S\delta_S ] A_{ij}. \eqno (3-25) $$
the anomalous  $S$ contribution ( see eq-(3-14) ) is  given by the last term  of (3-25) where the field-dependent $S$ parameters was  given in eq-(3-4). In this case one may use only $\phi_\mu$ for $\epsilon_S$ since there is the suitable $\gamma^\mu$ factor to contract vectorial indices with the $\phi_\mu$.   
The other covariant derivative : 

$$\gamma^\mu D^c_\mu \chi_{ij}  = \gamma^\mu ( D_\mu - (2+ {1\over 2} ) b_\mu ) \chi_{ij} - 
{1\over 2} \gamma^\mu \gamma^\nu D^c_\nu (A_{ij}) \psi_\mu - {1\over 2} \gamma^\mu F_{ij} \psi_\mu - 
(2 ) \gamma^\mu A_{ij} \phi_\mu + $$
$$ [  c_K\delta_K  \chi_{ij} ] . \eqno (3-26)$$
has an  explicit anomalous $K$-breaking pieces  in the last term  of the r.h.s of (3-26) and implicit 
$S$-breaking one  contained inside the ordinary derivative $D^c_\nu (A_{ij})$ as eq-(3-25) indicates. 
An example of the spurious  variations will be given next . 

As a result of the anomalous/spurious contributions to the ordinary derivatives the ordinary $Q$-transformations laws of the supermultiplet in eqs-(3-18, 3-19, 3-20) will acquire anomalous contributions themselves :

$$   D^c_\nu (A_{ij})|_{spurious} \Rightarrow             [ \delta_Q \chi_{ij} ]|_{spurious}  \sim  ({\bar \epsilon}\gamma^\mu  \phi_\mu ) (A_i\chi_j)    +...      \eqno(3-27)    $$

$$  D^c_\mu \chi_{ij}|_{spurious} \Rightarrow [ \delta_Q F_{ij} ]|_{spurious}  \sim   {\bar \epsilon}\gamma^\mu \gamma^\nu f^{m}_\mu e_{m\nu} A_i\chi_j +
{1\over 4}  ( { \bar \epsilon}\gamma^\mu  \phi_\mu )  ( {\bar \psi}_\nu  \gamma^\nu   A_i\chi_j ) + $$
$$ {1\over 2}   { \bar \epsilon}\gamma^\nu \gamma^\mu ( \partial_\mu + b_\mu ) (A_i A_j \phi_\nu )+...
\eqno (3-28)  $$

Therefore, under ordinary $Q$-variations the Lagrangian density (3-21) will acquire anomalous/spurious  pieces : 

$$\Delta|_{spurious}  = \delta_Q [ e L ]| _{sp}  =  e  [ (\delta_Q F_{ij})|_{sp}    +   {1\over 2} {\bar \psi_\mu} \gamma^\mu 
(\delta_Q \chi_{ij})|_{sp}  ]\not= 0. \eqno (3-29) $$ 
What the modified ${\tilde \delta_Q}{\cal S}  =[ \delta_Q + \delta' ] {\cal S}$ achieves is precisely to have 
$\delta' {\cal S} = -  \Delta|_{spurious}$ in order that the sum of the anomalous contributions cancels out giving zero : the $Q+K+S$ sum rule : $[ c_Q\delta_Q +c_K \delta_K + c_S \delta_S] {\cal S} = 0$.   

Notice that the first component $A_{ij} $ is $not$  subjected to he ${\cal Q}$-transformation as indicated by the first term of eq-(3-23). Also the members of the gravitational multiplet $ ( e^a_\mu, \psi_\mu, b_\mu) $ and the $\omega^{mn}_\mu, \phi_\mu, f^m_\mu$ transform properly under $S$ and $K$-transformations. Therefore, they will not be affected by the modified ${\cal Q}$-transformations. Since they do not have anomalous $S$ and $K$-contributions in their ordinary $Q$-transformations, there is no need to add a compensating $\delta' $ transformation.  

The modified covariant derivatives, appearing in the modified ${\cal Q}$-transformation laws (3-22,3-23, 3-24) of the supermultiplet (3-18, 3-19, 3-20 ) of Weyl weight $\omega = 2$  used to construct the action (3-21),  have now the appropriate form  :

$${\cal D} ^c_\mu (A_{ij}) = \partial_\mu A_{ij} - {1\over 2} {\bar \psi_\mu} \chi_{ij} - (2) b_\mu A_{ij} . 
\eqno (3-23) $$

$${\cal D} ^c_\mu \chi_{ij}  = ( D_\mu - (2+ {1\over 2} ) b_\mu ) \chi_{ij} - 
{1\over 2} \gamma^\nu {\cal D}^c_\nu (A_{ij}) \psi_\mu - {1\over 2} F_{ij} \psi_\mu - 
(2)  A_{ij} \phi_\mu . \eqno (3-24)$$

To sum up, after using a modified ${\cal Q}$-transformations (3-22)  to cancel out the anomalous $S,K$-transformations of the supermultiplet given by eqs- (3-18, 3-19, 3-20), used to construct the action (3-21), 
one will obtain an action invariant under the ${\cal Q}$-transformations ,  where one retains in the $modified$ covariant derivative ${\cal D}^c_\mu$ only the non-anomalous pieces and $discards$ the anomalous ones. The modified ${\tilde \delta_Q}$ transformations (3-22, 3-23, 3-24 )   achieves  precisely that.

Concluding, ${\tilde \delta_Q } {\cal S} = 0$ , the commutator of two modified ${\cal Q}$-transformations acting on the action is zero and can be written as a combination of $all$ the symmetries ( unbroken and broken) acting on the action. Because this sum is also zero this implies then that the combination, the $Q+K+S$ `` sum'' rule : 
$$ If~~~{\tilde\delta}_Q {\cal S} = 0 \Rightarrow  
[{\tilde \delta} ^{\epsilon_1}_Q, {\tilde\delta}^{\epsilon_2}_Q ] \int d^3x~  [ e {\cal L} ] =0 \Rightarrow   
{\bar \epsilon}^1_Q \gamma^\mu \epsilon^2_Q {\cal D}^c_\mu \int d^3x~ [ e {\cal L} ] = 0 \Rightarrow $$
$$\sum_A c_A \delta^{ \epsilon_A }_A 
\int d^3x~ [e {\cal L}]=  ( \sum_{unbroken} c_A \delta^{ \epsilon_A }_A [{\cal S}] )~  +    
[ c_Q \delta_Q^{\epsilon_Q} + c_S \delta^{ \epsilon_S }_S +  c_K\delta^{ \epsilon_K }_K ] [{\cal S}]  = 0. \eqno (3-6)$$
Since the unbroken symmetries yield a zero variation, from the last three terms of eq-(3-6) we can infer that the $combination$  of the anomalous $Q, S,K$ transformations acting on the action `` adds `` up to zero as required :

$$[c_Q \delta_Q + c_S\delta_S + c_K \delta_K]{\cal S} = 0 .  $$, 
despite the fact that individually   
$\delta_S {\cal S}  \not=0; \delta_K{\cal S} \not= 0 $ and $\delta_Q {\cal S}  \not=0$. The $K$-breaking of the action is crucial to compensate for the $Q$ and $S$-breaking pieces.

One can see why this procedure to write down actions soley invariant under the ${\cal Q}$-supersymmetry, translations, dilations and Lorentz, which form among themselves a $closed$ subalgebra, is very different from the standard  
Poincare supersymmetric actions  obtained from the full superconformal algebra by an explicit gauge fixing by hand of a suitable linear combination of the $Q,S$ supersymmetries and a gauging of the conformal boosts by setting $b_\mu = 0$. . 
It is the precise form of the action itself which $explicitly$ 
breaks down the $S$-supersymmetry,  and the conformal boosts symmetry,  leaving the ( modified) ${\cal Q} $ supersymmetry and the other symmetries intact. 
One has in essence a ${\cal Q}$-supersymmetric extension of the homothecy transformations, given by the invariance of the action under
${\cal Q}$ transformations, dilations, translations and Lorentz. 
We don't have $R$ terms, nor the explicit kinetic terms for the gravitino-field in our action; i.e there is no $f^m_\mu$ term.  
${\cal Q} $-supersymmetry is linearly realized after the elimination of $F^i$. No constraints arise like they did  
before in section {\bf 2}.  

To finalize we will show why the action is Weyl invariant despite the fact that there is no cancellation of $b_\mu$ terms. Due to the breakdown of conformal boost invariance, we should notice the explicit presence of the $b_\mu$ terms in 
the action resulting from the definition of 
the D'Alambert operator in eq-(3-7). This is tied up 
with the fact that there are no scalar curvature terms after one computes eqs-(3-18).

In $4D$ we learnt that the Weyl field 
$A_\mu$ decouples in the expression 
$$(D^{Weyl}_\mu D^{\mu}_{Weyl}  +\lambda R^{Weyl})\phi =0 $$
if one choosses the coupling $\lambda ={1\over 6}$ ( which must not be confused with the Weyl parameter associated with Weyl transformations). This is due to an exact cancellation between the 
$A_\mu$ field appearing in the Weyl scalar curvature and the D'Alambertian. It is no surprise that if 
the curvature is eliminated in eqs-(3-18 ) one will have $b_\mu$ terms remaining. However, Weyl covariance is maintained !
Eq-(3-7) that defines the three-dim version of the D'Alambert operator is explicitly Weyl covariant by $construction$. Notice the 
$explicit$ presence of the $b_\mu$ field in the $-[\omega (A)  -1]  b^\mu D^c_\mu A$ term in the 
r.h.s of (3-7) after unravelling the derivatives.  Under Weyl transformations these 
inhomogeneous pieces will cancel those stemming from the first term in the r.h.s of (3-7) 
$e^{-1}\partial_\nu (eg^{\mu\nu}D^c_\mu A)$. To have an explicit $b_\mu$ dependence does not imply that Weyl covariance
is destroyed. The definition in the l.h.s (3-7) of the D'Alambertian is explictly Weyl covariant. The tensor calculus does not break Weyl covariance. Eqs-(3-18) are defined explicitly in terms of well defined tensor products.

To finalize, we deemed  it important to add some final important remarks :

${\bullet}$   The action for the quadratic terms written above (3-21) $is ~ not $ the well known action for superconformal gravity that is separately invariant under $Q$ and $S$ supersymmetry . 
The acion above (3-21) is $not$  invariant under $S$ nor is invariant under conformal boosts as we have stated over and over again.  

${\bullet}$  The action (3-21) $is~ not$  the same as the well known superconformal action with the fermion bilinear  $S{\bar \chi}  \chi  $ term added .

${\bullet}$   The presence or absence of $S{\bar \chi}  \chi  $ is completely $irrelevant$  in the construction of the actions  :  
because the actions constructed  in eq-(3-21) 
are   $not$  the standard Poincare Supergravity action 
$without$ the fermion bilinear $S{\bar \chi}  \chi  $ term . The Poincare Supergravity action is $not$  separately invariant under $Q$ and $S$ supersymmeties. Once again, we wish to emphasize that it is only invariant under a suitable linear combination of $Q+S$ with a field dependent $\epsilon_S = - F\epsilon_Q$.

\bigskip 
\centerline{\bf 3.3  Supersymmetrization of the Quartic Derivative Terms}
\bigskip

The physical relevance of the quartic terms of the Skyrmion based action was discussed in {\bf 2.3} in relation to the physics of $D0$ branes in the large $N$ limit.     
The quartic derivative terms have the similar form of the usual terms, $F\wedge ^*F$,  in Yang Mills  theories. Also these quartic terms have a similar form to the 
Eguchi-Schild action for the string given by area-squared terms instead of the area action of the Dirac-Nambu-Goto strings. The Eguchi-Schld string is invariant under area-preserving reparametrizations only. 
These quartic terms in conjunction with the quadratic ones are necessary to reproduce the Dirac-Nambu-Goto action after the elimination of the three-metric via its algebraic equations of motion.

 We  proceed now to supersymmetrize the $L_4$ terms. Unless one introduces 
a coupling multiplying the quartic derivative bosonic terms, one cannot
obtain a superconformally invariant action (not even Q-invariant)  now because
these terms do not have the $net$ conformal weight of $\omega =2$ as  the kinetic
terms had. (We refer to the net weight of the first component of a multiplet so
that  $F$  has dimension three). For this reason we have to introduce the
following coupling function, a multiplet, that has no dynamical degrees of
freedom but which serves the purpose of rendering the quartic-derivative terms
with an overall dimension three to ensure that our action is in fact
dimensionless. We refrained from doing this sort of "trick" in the case of the
kinetic terms because such terms are devoid of a dimensional coupling constant.
The Dolan Tchrakian  action contains an arbitrary constant in front of the
quartic pieces and it is only the ratio between this constant and the
dimensionless constant in front of $L_2$ which is relevant. This constant must
have dimensions of $(length)^{3}$ since we have an extra piece of dimension
three stemming from the term ,$(\partial_\mu A)^2 $.

     Let us, then , introduce the coupling-function supermultiplet, 

$$\Sigma_0=(A_0;\chi_0;F_0).$$
whose Weyl weight is equal to $-3$  so that the tensor product of $\Sigma_0$ with
the following multiplet, to be defined below, has a conformal weight ,
$\omega=2$ as it is required in order to have Q-invariant actions.

Lets introduce the following multiplet ( with an overall $\omega =5$ so the $F$ terms have dimension six). 
$$K^{ijkl}_{\mu\nu\rho\tau} =K(\Sigma^i_\mu;\Sigma^j_\nu)
\otimes{T[K(\Sigma^k_\rho;\Sigma^l_\tau)}] +({ij\leftrightarrow kl})~
and~({\mu\nu\leftrightarrow\rho\tau})-$$
$$T[K(\Sigma^i_\mu;\Sigma^j_\nu)\otimes
{K(\Sigma^k_\rho;\Sigma^l_\tau)}]. \eqno (3-30a)$$
                  
$$K(\Sigma ,\Sigma)=\Sigma^i_C \otimes T_C(\Sigma^j_C ) + T_C(\Sigma^i_C
)\otimes \Sigma^j_C \equiv (A^K_{ij};~\chi^K_{ij}; ~F^K_{ij})  \eqno (3-30b)$$
when we construct the action one must perform the permutation of the spacetime indices and contract, afterwards at the end of the calculation, all spacetime indices with $\eta_{ij}\eta_{kl}$. 

This combination of supermultiplets is the adequate one to retrieve (2-7) at the bosonic level and also the
one which ensures that the $R$ terms do cancel from the final
answer. This is a similar case like the construction of the quadratic terms which was devoid of $R$ and kinetic gravitino terms, as shown explicitly in eqs-(3-18,3-19, 3-20). 

However, we must be sure that indeed there are no $R$ and kinetic gravitino terms $resurfacing$ again due 
to any potential anomalous $S,K$-transformations encoded inside the covariant derivatives, that are associated with the $T_C$ operation.. For example, we had $f^m_\mu$ terms emerging due to the spurious terms contained inside $D^c_\mu \chi_{ij}$ given by eq-(3-27); i.e there may be anomalous $S$ and $K$-transformations resurfacing  again in the derivatives, D'Alambertian, etc...which $define$ such quartic multiplet. 

For this reason one $must ~ not$ include the anomalous $S,K$ kinetic multiplet 
$T_C( \Sigma^i_C \otimes \Sigma^j_C)$ in the definition of the $K_c(\Sigma_c ,\Sigma_c)$ appearing in eqs-(3-30). $K_c(\Sigma_c ,\Sigma_c)$ as defined by eqs-(3-30) has the correct $S,K$ transformations that allowed us to write down the fully invariant superconformally kinetic action in $D =3 $ given by eq-(3-15, 3-16).  
The Tensor calculus is defined for those supermutiplet with well behaved $S,K$ transformations laws. It is meaningless to apply the tensor calculus for those objects which have spurious $S,K$ transformations laws. Therefore, the $entries~ inside$ the $T_C$ operations in (3-30a, 3-30b) must be well behaved supermultiplets.  

A similar calculation yields the components of the supersymmetric-quartic-derivative
terms: 

$$A_{ijkl}= { \bar\chi}^K_{ij}   \chi^K_{kl}.\eqno(3-31a)$$                                  

$$\chi_{ijkl} =F^K_{ij}\chi^K_{kl} +F^K_{kl}\chi^K_{ij} +A^K_{ij}\gamma^\mu D^c_\mu (\omega ={2})\chi^K_{kl}  
+A^K_{kl}\gamma^\mu D^c_\mu (\omega ={2})\chi^K_{ij}-$$
$$\gamma^\mu D^c_\mu (\omega =4)[A^K_{ij}\chi^K_{kl} + A^K_{kl}\chi^K_{ij}].        
\eqno(3-31b)$$            

$$F_{ijkl}= A^K_{ij} \triangle ( \omega = 2)  A^K_{kl}  +A^K_{kl}  \triangle (\omega = 2 )  A^K_{ij}   
+2F^K_{ij} F^K_{kl} 
-{\bar \chi}^K_{ij}  \gamma^\mu D^c_\mu ( \omega ={2})\chi^K_{kl}  -$$      
$$  {\bar \chi}^K_{kl}  \gamma^\mu D^c_\mu ( \omega = 2)\chi^K_{ij}  
- \triangle ( \omega = 4)[A^K_{ij}A^K_{kl}]. \eqno (3-31c)$$       
Where we have used the abreviations $A^K_{ij},\chi^K_{ij}$ and
$F^K_{ij}$ given by the well behaved kinetic multiplet of eq-(3-30b). The derivatives acting on composite fields must appear with the right Weyl weight. 

Notice the $similarity$, as one should have,  between eqs-(3-18, 3-19, 3-20) used to construct the quadratic 
actions in (3-21) and eqs-(3-31) required to construct the supersymmetric 
quartic derivative terms , both in
$form$  and in the $values$ of the coefficients. This it due to the tensor calculus nature of the 
supermultiplets. This is a sign of consistency  and should serve as a check. Such similarity will be important to obtain the equations of motion associated with the quartic terms. 

Because we have used the proper components $A^K_{ij}; \chi^K_{ij}, F^K_{ij} $ associated with the definition of $K_c(\Sigma_c\otimes \Sigma_c)$ given by eqs-(3-30b) we are now certain that $no$ spurious $R$ and kinetic gravitino terms will re-emerge due to any potential anomalous $S,K$ contributions encoded inside the 
supercovariant derivatives used to define the  ( well behaved ) kinetic $K_c(\Sigma_c\otimes \Sigma_c)$ 
given by eq-(3-30b); i.e the entries inside the $T_C$ operations of (3-30a) have well behaved $S,K$ transformations. 
However, we are not implying the the quartic multiplet eq-(3-30a) will have well defined $Q$ transformations !. Such quartic multiplet will break $S,K$ symmetries like it occured before with the quadratic multiplet of eqs-(3-18, 3-19).  

Hence , despite the fact that the defining quartic multiplet (3-30a) is devoid of curvature terms we are once again faced with the anomalous $Q$ transformations of the action due to the spurious $S,K$-variations which will appear in the ordinary $Q$ variations of the action using the ordinary supercovariant derivatives. To cure this anomalous variations of the action we proceed exactly as before by using a modified ${\cal Q}$ transformations that will cancel the spurious variations. 

Finally  we take the tensor product of the latter multiplet given above in eq-(3-28) and the
coupling-function multiplet to have the right dimensions :
$$\Sigma_0\otimes(A^{ijkl};\chi^{ijkl};F^{ijkl})=(A_0A^{ijkl};A_0\chi^{ijkl}+\chi_0A^{ijkl};
A_0F^{ijkl}+F_0A^{ijkl} -{\bar \chi_0}\chi^{ijkl}). \eqno (3-32)$$
It is important to emphasize that the coupling function multiplet, $\Sigma_0$ is $not$ a Lagrange multiplier. A variation of the action w.r.t the 
coupling function multiplet is $not$ the correct way to determine its expression in terms of the spinning membrane matter fields. As it was mentioned in the introduction, the way to determine their  values is through the algebraic elimination of the members of the gravitational supermultiplet supplemented by the embedding condition : $D^c_\mu A_0=0$. We will discuss this issue further in Appendix {\bf A} and in the next section.   

The complete ${\cal Q} $ supersymmetric extension of $L_4$ requires adding terms
which result as permutations of ${ijkl\leftrightarrow ilkj\leftrightarrow
kjil\leftrightarrow klij}$ keeping $\eta_{ij}\eta_{kl}$ fixed and inserting (3-31, 3-32) into the Lagrangian density of the same form given by (3-15) with the only difference that now one uses the components 
$({\cal A}_{ijkl}; \chi_{ijkl}; {\cal F}_{ijkl})$.

The action corresponding to the quartic derivative terms is :

$${\cal L}_4= e\eta_{ij}\eta_{kl}~[A_0F^{ijkl}+F_0A^{ijkl} -{\bar \chi_0}\chi^{ijkl}+{1\over 2}{\bar \psi}_\mu \gamma^\mu (A_0\chi^{ijkl}+\chi_0A^{ijkl})
+{1\over 2}A_0A^{ijkl}{\bar \psi}_\mu \sigma^{\mu\nu} \psi_\nu] . \eqno (3-33)$$
One must include permutations ${ijkl\leftrightarrow ilkj\leftrightarrow
kjil\leftrightarrow klij}$ with the appropriate signs while keeping         
$\eta_{ij}\eta_{kl}$ fixed. The composite fields $A_{ijkl}, \chi_{ijkl}, F_{ijkl}$ 
are given in terms of  eqs-(3-32) and eqs-(3-31).

The action will be invariant under the $modified$  ${\cal Q}$-supersymmetry transformations laws, 
which have the same structure as eqs-(3-23) :

$${\tilde \delta_Q} {\cal A}_{ijkl} = {\bar \epsilon}  \chi_{ijkl}.~~~{\tilde \delta_Q}\chi_{ijkl}= 
F_{ijkl} \epsilon +  \gamma^\mu {\cal D}^c_\mu {\cal A}_{ijkl}\epsilon.~~~{\tilde \delta_Q}F_{ijkl}= 
{\bar \epsilon}   \gamma^\mu    {\cal D}^c_\mu \chi_{ijkl}. \eqno (3-34)$$

with : 

$${\cal D} ^c_\mu (A_{ijkl}) = \partial_\mu A_{ijkl} - {1\over 2} {\bar \psi_\mu} \chi_{ijkl} - 
(2) b_\mu A_{ijkl} . \eqno (3-35a) $$

$${\cal D} ^c_\mu \chi_{ijkl}  = ( D_\mu - (2+ {1\over 2} ) b_\mu ) \chi_{ijkl} - 
{1\over 2} \gamma^\nu {\cal D}^c_\nu A_{ijkl} \psi_\mu - {1\over 2} F_{ijkl} \psi_\mu - 
(2)  A_{ijkl} \phi_\mu . \eqno (3-35b )$$

The quartic multiplet will then have the well behaved modified ${\cal Q}$ transformation laws devoid of spurious terms resulting from anomalous $S,K$ variations. The gauge algebra will then close as required. 
One finally has :

(i) Linearly realized modified  ${\cal Q}$ supersymmetry.
(ii) Absence of $R$ terms and explicit absence of the gravitino kinetic terms.
(iii) A polynomial Lagrangian in the fields. (iv) Closure of two succesive modified ${\cal Q}$-transformationson the action. 
(v) The gauge fields are not subjected to the modified  ${\cal Q}$-transformations since they transform properly under $S$ and $K$-transformations. Therefore, the  gauge algebra closes on them.

\bigskip

\bigskip
\centerline{\bf 3.4  Elimination of the Auxiliary fields} 
\smallskip
The auxiliary field $F$ appearing in the quadratic action, for example,  has the desired quadratic $F^2$ pieces  that allows it to be 
eliminated algebraically without introducing constraints among the $A,\chi$ fields. Upon the elimination of $F$ yields $F=F(A,\chi)$. 
Lets us mention now why in the usual constructions, one must do things in the right order . For example, if one were to fix the Weyl invariance in the ordinary superconformally invariant actions, $after$ having eliminated the auxiliary fields,  one has in the last 
relation that by setting $F=  F(A,\chi)= constant$,  one will reintroduce constraints; 
however these are $not$ entirely due to the algebraic elimination of the auxiliary field, $F$, but to 
the fact that a gauge condition has been selected $after$ the auxiliary fields were eliminated.  A gauge choice naturally constrains fields or some of its components. For this reason, in the ordinary constructions of 
invariant actions one fixes $first$ the superfluous symmetries and afterwards one eliminates auxilary fields.

However, the construction presented in this work is very $different$  from the standard ones, 
because we are already $starting$ from a Weyl covariantized action : the Weyl symmetry is no longer a 
$superfluous$ symmetry  to be fixed $prior$ to the elimination of the auxiliary fields. 
On the contrary : one firstly eliminates the auxiliary fields in the Weyl-invariant ${\cal Q} $-supersymmetric membrane actions, and then one sets the fermions to zero to recover, finally,  the initial Weyl invariant action we started with : 
the Weyl-covariantized Dolan-Tchrakian action.

In this subsection we will discuss in detail the variation of the auxiliary fields corresponding to the membrane matter fields. 
The equations of motion of the quadratic and quartic terms w.r.t the $A_i,\chi_i, F_i$ fields are what one needs to show that no problems arise upon elimination of the auxiliary fields.   
Eliminating the auxiliary fields, ${\partial L/{\partial F^i}}=0$,
and setting the Fermi fields to zero we must recover the orginal Weyl-covariantized Dolan-Tchrakian 
action (WCDT). Furthermore, the order in which we perform this should yield identical
results: set the Fermi fields to zero and eliminate the auxiliary fields and vice versa.

In the past there has been a lot of debate concerning the elimination of the auxiliary fields, $F^i,F_0$ [8]. 
We have shown that there are $no$ constraints ( which will spoil the linear realization of supersymmetry in the Hilbert space of physical fields)  among the physical membrane fields after the elimination of the auxiliary fields $F^i$. 
Earlier in [8] we discussed what happens upon the elimination of $F_0$. Since supersymmetry rotates 
field equations into field equations among given  members of a supermultiplet, one $must$ and $should$ 
eliminate as well the remaining members of the coupling supermultiplet , $A_0,\chi_0$. When this is done the $full$ 
quartic supersymmetric terms are constrained to zero. This does $not$ imply that the quartic terms in the bosonic sector are zero; it is the 
whole sum of the bosonic quartic terms and their supersymmetrization   which is constrained to zero. 
This constraint is $unnatural$ because the coupling multiplet must $not$ be treated as a Lagrange multiplier. Secondly,  upon setting the fermions to zero the 
auxiliary field $F_0$ decouples from the action any way [8] and , thirdly, if one varies the couplings while 
maintaining the embedding condition, $D^c_\mu A_0 =0$,   one $cannot$ longer vary  simultaneously the 
members of the 
gravitational super multiplet as one should in order to retrieve the Dirac-Nambu-Goto actions. Because if 
this is done too many 
restrictions will arise among the three supermultiplets.

The correct procedure ( instead of the one in [8]) , due to the fact that the $b_\mu$ field 
does $not$ decouple from the action and also because it is a member of the supermultiplet 
containg the graviton and gravitino, is to vary those members of the gravitational supermultiplet and $not$ those  
of the coupling function supermultiplet. 
The superconformal covariant embedding condition, $D^c_\mu A_0 =0$, plus its series of supersymmetric transformations  
( ``rotations'') , will furnish the desired relations between the 
fields $(A_0,\chi_0,F_0)$ 
and the $3D$ background gravitational supermultiplet $(e^m_\mu, \psi_\mu, b_\mu)$ which,  in turn , can be 
determined in terms of the physical membrane fields,  $(A^i, \chi^i, F^i)$  resulting from  the $algebraic$ 
elimination of the gravitational supermultiplet ( non-propagating field equations)  upon the variation of the action. This is shown in Appendix {\bf A}.

Eliminating the $F^i$ field yields the value for $F^j$=0, after
we set the Fermi fields to zero. Conversely, setting the fermions to zero and eliminating the 
$F^i$ fields yields zero as the viable solution for the $F^i$ fields since we don't wish to 
generate constraints
among our physical fields. It is precisely when we set the Fermions to zero that the
$F_0A^{ijkl}$ term vanishes and the $\bar {\chi_0}  \chi^{ijkl}$ terms as well. We
are left only with the $A_0F^{ijkl}$ piece belonging to the bosonic quartic terms in the WCDT action as 
intended. 
Finally, fixing the Weyl gauge invariance by setting $A_0=g$ renders the WCDT action in the original DT form once the 
embedding condition $D^{Weyl}_\mu A_0 =0 $ is implemented ( once the fermions are set to zero the 
superconformal derivative becomes the ordinary Weyl derivative). $A_0=g \Rightarrow b_\mu \sim \partial_\mu lnA_0=0$.

    It is fairly clear that we have enforced ${\cal Q}$-supersymmetry. The fields which
comprised the "coupling" function do not explictly break conformal invariance; on the contary, conformal invariance demands their introduction to 
accomodate dimensions properly.  One does $not$ vary couplings in the actions we are familiar with. Varying the members of the coupling function supermultiplet 
constrains, naturally,  the supersymmetric quartic derivative terms to zero as discussed in the introduction. This is due to the fact that all the members of the 
coupling function supermultiplet appear as a $multiplicative$ factor. Since the members of the 
coupling function supermultiplet are $not$ Lagrange multipliers enforcing any constraints to vary these multiplicative couplings is physically $meaningless$. A question immediately arises : If one does not vary the couplings in the action how then can one determine their values ? 

This was thoroughly explained earlier in the introduction. Fistly, one eliminates algebraically the members of the gravitational supermultiplet via their variation in the action ( the gravitational supermultiplet is not dynamical ) and, afterwards, one imposes the embedding condition : $D^c_\mu A_0=0$. Upon supersymmetry transformations, the remaining members of the coupling supermultiplet are eliminated as well.  In this way one will be able to express the couplings and the members of the gravitational supermultiplet in terms of the membrane matter fields and their superpartners. Further details of how this is attained are discussed in the Appendix A. In particular, we will discuss in detail how the embedding condition   $D^c_\mu A_0=0$ is indeed compatible with the membrane matter fields equations of motion associated with the WCDT action. As a reminder, the embedding condition is just the statement that the coupling function $A_0$ is a "constant" from the ( super) Weyl poin!
t of view. We believe that there is no need to repeat this discussion again.

Concluding this section, we mantained  $conformal$ 
invariance in the sought-after ${\cal Q}$ spinning membrane and obtained 
a Weyl invariant and a ${\cal Q} $-spinning polynomial membrane action 
in contradistinction with the Lindstrom-Rocek's  result of a Weyl-invariant Poincare supersymmetric non-polynomial action.   
This was the crux of Weyl-covariantizing  the DT action 
and constructing afterwards a  superconformally invariant action.  
We emphasize once again that we have supersymmetrized 
the Weyl covariant version of the DT action and $not$ the DT action.

\bigskip
\centerline {\bf 4. Concluding Remarks : Quantization and Anomalies}
\bigskip

Concluding, we mantained  conformal 
invariance and constructed  a modified ${\cal Q}$ supersymmetry, a $Q+K+S$ `` sum `` rule, to ensure closure of the gauge algebra in the sought-after ${\cal Q}$ spinning membrane : a Weyl invariant and ${\cal Q}$ spinning polynomial membrane action which is invariant under the supersymmetric extension of the homothecy transformations. This is very $different$   
than the Lindstrom-Rocek construction of a Weyl-invariant Poincare supersymmetric non-polynomial action. 
We have finally ${\cal Q}$ supersymmetrized the Weyl covariantized  Dolan-Tchrakian action. The kinetic terms and quartic terms are ${\cal Q}$-invariant by construction. The latter
ones were ${\cal Q}$-invariant with the aid of an extra multiplet, the
"coupling" function multiplet whose weight is precisely equal to $-3$ to ensure
that our action is dimensionless and scale invariant. After 
eliminating the $F^i$ auxiliary fields, having set the
Fermi fields to zero, 
and fixing the dilational invariance 
we retrieve the Dolan-Tchrakian Lagrangian for the membrane ( once the embedding condition is implemented). 

The main
point of this paper is to show that one can have a ${\cal Q}$-spinning membrane 
solely if we wish to satisfy all of the requirements listed in the introduction. $"Q+S"$  invariance can only be implemented in non-polynomial actions as Rocek and
Lindstrom showed [5]. For more recent actions for super $p$-branes see [26] and for world volume supersymmetric Born Infeld actions see [27].

Since the only obstruction to fixing the gauge $b_\mu\sim \partial_\mu lnA_0 =0$ globally is topological
it is warranted to study the topological behaviour of these 3-dim gauge fields
and see what connections these may have with Witten's Topological QFT, 
Chern-Simmons 3-dim Gravity and with other non-perturbative phenomena in three
dimensions [7].

It is desirable to repeat the anomaly cancellation analysis for the spinning membrane along the same lines as it was done for the spinning string. Marquard, Kaiser and Scholl [19] 
have shown that the lightcone gauge Lorentz algebra closes for the bosonic membrane in 
$D=27$ dimensions and in $D=11$ for the supermembrane case. This involved writing the Noether charges associated with the global target space Poincare ( super) symmetries  
and computing the light-cone gauge Lorentz algebra commutators. 

In general, even before the evaluation of the Lorentz algebra commutators, the construction of the
quantum Lorentz generators requires extra renormalizations as it is the case for composite operators in QCD [21]. Marquard et al [19] used the Operator Product Expansion and a suitable regularization method to compute the quantum Lorentz generators and the light-cone Lorentz algebra commutators. 
The extra terms induced by the operator ordering ambiguities
drop out of the light cone Lorentz algebra in the critical dimensions only; i.e the Lorentz algebra closes in the $D=11$ dimensions.  
In the ( super) string case is not necessary to follow this construction since the theory reduces to that of a two-dim $free$ field theory. Thus one can invoke 
to the  standard mode expansion of the free fields and recur to the usual normal ordering procedure to solve operator ordering ambiguities and compute the Lorentz algebra commutators. 

The lightcone gauge condition on the superstring coordinates $X, \theta$ is :
$X^+=x^+_0 +\tau.~~~\Gamma^+ \theta =0. $ 
Using the Virasoro constraints, resulting from the two-dim metric equations of motion,  and the lightcone gauge condition allows to write all the other componets of the fields ( longitudinal)   in terms of the transverse degrees of freedom  only.    
The Lorentz generators :
$J^{\mu\nu}=M^{\mu\nu}+K^{\mu\nu}$ containing the bosonic and fermionic pieces
can then be written in terms of the the transverse fields and their 
bosonic/fermionic oscillators
mode expansion. The relevant commutator is the one involving the boosts 
operators :$[J^{i-}, J^{j-}]=0 $
from which the critical dimension and the Regge intercept is obtained. 

Simlar considerations apply in the spinning string case. One starts with the light cone gauge condition :
$X^+=x^+_0 +\tau.~\psi^+  =0$
Under supersymmetry transformations this gauge is preserved :
$\delta_Q X^+= {\bar \epsilon }\psi^+=0$ and no addition of compensating gauge terms is   required. 
The use of the super-Virasoro constraints, resulting from the two-metric and gravitino field equations, allows to express the Lorentz Noether charges in terms of the mode expansions of the transverse field components only. 
The commutator $[J^{i-}, J^{j-}]=0$ will again select $D=10$ as the critical dimension. 

The spinning membrane is far more complicated mainly because the theory is highly nonlinear and no simple mode expansion can be obtained in general. An Operator Product Expansion and the regularization method like the one employed in 
[19] is perhaps the most promising avenue to calculate the quantum Lorentz generators and the light cone Lorentz algebra commutators. 

However, the action written here in eqs-(3-13, 3-18) involving quadratic and quartic terms is already very complicated to start with. To evaluate, firstly,  the Noether charges associated with the global target space Lorentz symmetries requires suitable renormalizations, let alone to compute the OPE of the currents in the 
light cone Lorentz algebra commutators . The final step would be to show that 
the terms induced by operator ordering amibiguities cancel in the light cone Lorentz algebra only in the critical dimension, $D=11$. 
This is a very complicated task. 

Due to the $similarities$ between our explicitly Lorentz covariant Skyrmion-based action (2-14a, 2-14b) and the light-cone supermembrane action,  one could attempt to follow the steps taken by several authors in the literature :  The computation of the Lorentz symmetry of the $D=11$ supermembrane in the light cone gauge was completed by [23] following the program initiated by [24]. An earlier unpublished proof of the Lorentz symmetry of the light cone eleven dimensional supermembrane
 was given by [22]. Another approach has relied on the equivalence between the light cone eleven dimensional supermembrane and large $N$ $SU(N)$ matrix models. Within this context of a $SU(N)$ supersymmetric gauge quantum mechanical model , after the group of area preserving diffs is suitably regulated, it was shown that the Lorentz algebra closes to leading nontrivial order at large $N$ by [20]; i.e the lowest non-trivial terms induced by quantum mechanical operator ordering do cancel. Strangely enough, this calculation seems to work for all classically allowed dimensions so there was no indication of a critical dimension $D=11$ at the order considered [21]. 

Again, the spinning membrane is another story. There is no obvious equivalence between the spinning membrane spectrum and that of the supermembrane. In the string case, the 
superstring  and the spinning  string formulation were physically equivalent 
and anomaly free in $D=10$. This result was based on the role of bosonization  in two dimensions, the GSO proyection of the NSR spinning string  to ensure that the spacetime spectrum was indeed supersymmetric and the triality condition of $SO(8)$. In the membrane case, a full fledged quantization has not been achieved yet except in some special simpler cases where a semiclassical quantization scheme was attained ; the issue of massless states and the existence of a normalizable ground state has not been settled yet. Furthermore, it has been argued that the membrane theory is a second quantized theory from the very beginning and the very notion of a one or 
multiparticle states can only be extracted in certain asymptotic regions if it makes sense at all [21].   

We have been concentrating solely on the Lorentz invariance property and whether or not one can extract the critical dimension $D=11$ for the spinning membrane and the supermembrane. There is also the issue of area-preserving diffs anomalies and fermionic kappa symmetries  ( in the supermembrane case) .  

Concerning the  area-preserving diffs anomalies, we were able to show [25] that the expected critical dimensions for the bosonic membrane and the supermembrane ( $D=27, D=11$) were related to the number of spacetime dimensions of an anomaly-free non-critical (super) $W_\infty$ string. $W_\infty$ algebra is the area-preserving diffs of a plane. $W_{1+\infty}$ is that of a cylinder. A BRST
analysis revealed that a very special sector of the ( self dual ) membrane spectrum 
has  a relationship to a first unitary minimal model of a 
( super) $W_N$ algebra $adjoined$ to a critical ( super) $W_N$ string spectrum in the $N=\infty$ limit. In this sense, the self dual ( super) membrane, moving in flat target spacetime backgrounds, admits non-critical $W_\infty$ ( super) strings in its spectrum and is anomaly free in $D=27, D=11$ dimensions , respectively.  

To quantize our bosonic Skyrmion-based action (2-14a, 2-14b) and its ${\cal Q}$ supersymmetric action is a challenging enterprise before we can begin to answer the question of anomalies and critical dimensions.   
A covariant BRST-Batalin Vilkovski formalism and what we know about the quantum field theory of Skyrmions could be a starting point. The regularization procedure and the use of the OPE method of [19] is another viable, but equally difficult, avenue.

For other constructions of extended objects using a modifed measure we refer to Guendelmann [30]. It will be very interesting to see if one could write a spinning version of Guendelmann's actions. For the WKB quantum equivalence of diverse $p$-brane actions we refer to [31].

\centerline {\bf Acknowledgements}
\medskip
We are indebted to the Center for Theoretical Studies of Physical Systems for support and to Luis Boya for
discussions at the University of Zaragoza, Spain. Also to Yuval Ne'eman for having suggested this problem and 
George Sudarshan for his help. We are also indebted to Eduardo Guendelman , Euro Spallucci, Alex Granik for recent discussions.

\medskip

\bigskip
\centerline{\bf APPENDIX A } 
\bigskip
We are going to show that the embedding condition $D^c_\mu A_0=0$, which states simply that the coupling function $A_0$ is `` constant'' from the superconformal point of view, 
is compatible with the membrane matter fields equations of motion associated with the WCDT action.  Also we will show how the members of the gravitational supermultiplet and the coupling function supermultiplet can be expressed in terms of the membrane matter fields.

Such embedding condition is written in an explicit superconformally covariant fashion. A supersymmetry transformation will ``rotate '' the latter embedding condition imposing additional conditions on the remaing members of the coupling function multiplet. Since all the members of the coupling function multiplet appear as a $multiplicative$ factor on the quartic terms as shown in the text , without loss of generality , as an example, we are going to concentrate on the particular case below. 
The argument displayed by the equations of Appendix {\bf A}, below,  can be generalized to the full supersymmetric case. 

A rough illustration of the type of terms to be studied after eliminating the auxiliary $F^i$ fields and 
setting the fermions to zero 
is the following. 
The Weyl covariantized DT action 
has the form ( we are not including the target spacetime indices nor the antisymmetrization 
of the $3D$ indices as well):
$$-(\partial_\mu A -\omega (A)  b_\mu A)^2 +A_0(\partial_\mu A -\omega (A)  b_\mu
A)^4.\eqno (A-1) $$
where $A_0 <0$ since from (2-7) we know that the relative sign between the quadratic and quartic terms 
is the same. Eliminating the $b_\mu$ after a variation w.r.t the $b_\mu$ 
yields:
$$  (D_\mu^{Weyl} A) +2(-A_0)(D_\mu^{Weyl} A )^3 =0. \eqno (A-2) $$
where we have factored  out the term $-\omega (A)  A $ (which should not be
constrained to zero). therefore
one gets two possible solutions:
$$ 1-2A_0(D_\mu ^{Weyl} A )^2 =0 \Rightarrow (D_\mu ^{Weyl} A )^2 ={1\over 2 A_0}<0. \eqno (A-3) $$
this is consistent with the timelike condition of the vector $(D_\mu ^{Weyl} A$). 
The other condition is 
$(D_\mu ^{Weyl} A )=0$ which is unacceptable because it  
constrains the action to zero which is not very physical  whereas the former condition is fine. 
Implementing the    
embedding condition $(D_\mu ^{Weyl} A_0 )=0\Rightarrow b_\mu \sim \partial_\mu ln A_0$ in (A-3) yields
 the 
desired relationship among $A_0,A,b_\mu$ :

$$(D_\mu ^{Weyl} A )^2 = [\partial_\mu A - \omega (A)  (-{1\over 3} \partial_\mu ln A_0) A]^2 =
{1\over 2 A_0}. \eqno (A-4)$$
The last equation yields the relation between $A_0$ and  $A$ and the relation $b_\mu =-{1\over 3} (\partial_\mu ln A_0)$
determines $b_\mu$ in terms of $A$. In this fashion one has found the relationship among all the fields in terms of the physical membrane coordinates. It is true that (A-4) is not an algebraic relation 
between $A_0,A$, however this does not spoil the linear realization of supersymmetry among the membrane's fields. 

The construction in section III presupposes the fact that one can find a gauge where
(simultaneously) the scalar coupling field 
$A_0$  can be gauged to a constant and the
$ b_\mu$ field to zero. We will show now that the condition  $D_\mu^{Weyl} A_0 =0$ is indeed compatible with the membrane matter fields equations of motion associated with 
the WCDT action. The equations of motion of the $A$ fields stemming from the quartic derivative 
terms of the WCDT action are of the form :

$$D^{Weyl}_\mu [{\delta S^{bosonic}_4 \over \delta (D^{Weyl}_\mu A)}] 
\sim (D_\mu ^{Weyl} A_0 )(D_\mu ^{Weyl} A )^3
+A_0 D^{Weyl}_\mu [(D_\mu ^{Weyl} A )^3]. \eqno (A-5)$$

The above expression is Weyl covariant and we have assumed that there are no boundary terms in our action and that the fields vanish fast enough at infinity....As it is usual in these variational problems we have integrated by parts and generalized Stokes law to the Weyl space. 
Now, if, and only if, the equations of motion ( of the matter fields ) of the WCDT action are indeed the Weyl covariant extension of the equations of motion ( of the matter fields)  of the DT action then the condition 
$(D_\mu ^{Weyl} A_0 )=0$ is indeed compatible and consistent . 
This follows immediately since there is no analog of those terms in the DT equations of motion. The second terms in the r.h.s of (A-5) are the only ones consistent with the 
Weyl covariant extension on the DT matter fields equations of motion. This implies that the first terms in the r.h.s must be zero and hence $(D^{Weyl}_\mu A_0) =0$.

A similar argument follows in the full fledged superconformal covariant case by replacing the Weyl covariant 
derivative by the superconformal one :  $(D_\mu ^c A_0 )=0$. This is due to the fact that all the members of the coupling function multiplet appear as a 
$multiplicative$ factor.
Hence, one will have in the equations of motion variations involving the superconformally covariant derivative of the form :
$$D^c_\mu [{\delta S_4 \over \delta (D^c_\mu A)}] =  (D^c_\mu A_0)(....)+A_0 (....)+...\eqno (A-6)$$       
If, and only if, the matter fields equations of motion are indeed the superconformally covariant extension of the matter fields equations of motion of the DT action,  it follows then that $(D_\mu ^c A_0 )=0$ is indeed a compatible embedding condition. This is a result that follows from the fact that the first terms in the r.h.s of (A-6) are not consistent with the superconformaly covariant extension of the DT matter fields equations of motion. Hence,  $(D_\mu ^c A_0 )=0$ follows. 
In the full supersymmetric case,  supersymmetry transformations   will
``rotate'' the embedding condition yielding the additional conditions on the 
remaining fields $\chi_0, F_0$.

To finalize this Appendix we point out that the only obstruction in setting
$b_\mu$ to zero must be topological in origin. We saw in section II
that it was the elimination of S which originated the constraint
$\bar{\chi}\chi$ =0. Such S term had the same form as an effective mass resulting from a 
fermion-condensate.
Whereas here, upon the "trade-off" $ b_\mu \leftrightarrow {{1\over4}\gamma_\mu S}
$ we may  encounter topological obstructions in setting $b_\mu$ =0 gobally
and, henceforth, in Q-supersymmetrizing the Dolan-Tchrakian action; i.e.  to
obtain the exact bosonic limit from the Q-supersymmetric action. 
\medskip

\centerline {\bf References}
\smallskip
1-.P.S Howe and  Tucker, J. Math. Physics {\bf 19} (4) (1978) 869.

J.Math. Physics {\bf 19}  (5) (1978) 981. J. Physics {\bf A 10} (9) (1977) L 155.

2-.M.Duff,  "Supermembranes, The first 15 Weeks ". CERN-TH- 4797 (1987).

Class. Quan. Grav {\bf 6} (1989) 1577. 

3-.E.Bergshoeff, E.Sezgin and P.K.Townsend,  Phys.  Lett.  {\bf B 209} (1988) 451.

4-.T.Uematsu,  Z.Physics {\bf C 29}  (1985) 143-146 and {\bf C32}  (1986) 33-42.

5-. U.Lindstrom,  M. Rocek,  Phys. Letters {\bf B 218}  (1988) 207.

6-.B.P.Dolan,  D.H.Tchrakian,  Physics Letters {\bf B 198}  (1987) 447.

7-S.Deser, R.Jackiw and  S. Templeton: Ann. Physics {\bf 140} (1982) 372.

8-C. Castro, Journal of Group Theory in Physics {\bf 1} (2) (1993) 215.   

J. Chaos, Solitons and Fractals, {\bf 7} (5) (1996) 711.

9-J. Polchinski , TASI lectures on D-branes, hep-th/9611050. 

J. Polchinski , S. Chaudhuri and C. Johnson : `` Notes on D-branes, hep-th/9602052. 

10-E. Witten, Nucl. Phys. {\bf B 443}  (1995) 85.

C. Hull and P. Townsend, Nucl. Phys. {\bf B 438} (1995) 109.

C. Vafa, `` Evidence for F theory `` , hep-th/9602022. 

J.Schwarz : `` The Power of M theory'' , hep-th/9510086

R. Dijkgraaf : `` Les Houches Lectures on Fields , Strings and Duality ``,  

hep-th/9703136

11- T.H.R.  Skyrme , Proc. Roy. Soc. {\bf A 260} (1961) 127. 

C. Houghton, N. Manton, P. Sutcliffe, `` Rational Maps, Monopoles and Skyrmions ``.

hep-th/9705151. 

N. Manton, Comm. Math. Phys {\bf 111} (1987) 469. 

M. Duff, S. Deser and C. Isham, Nucl. Phys. {\bf B 114}  (1976) 29.

12- T. Banks, W. Fischler, S. Shenker and L. Susskind, 

`` M theory as a Matrix model, a Conjecture `` hep-th/9610043. 

Phys. Rev {\bf D 55} (1997) 5112.

13- W. Taylor : `` Lectures on $D$- branes, Gauge Theory and M(atrices)'' 

hep-th/9801182. 

14-   C. Zachos, D. Fairlie and T. Curtright : `` Matrix Membranes and Integrability `` 

hep-th/ 9709042.

15-Y. Ne'eman , E. Eizenberg : `` Membranes and Other Extendons ( p-branes) `` 

World Scientific Lecture Notes in Physics {\bf vol 39} . (1995).

16- Ivanova, Popov : Jour. Math. Phys. {\bf 34} (2) (1993) 674. 

Theor. Math. Phys. {\bf 94} (2) (1993) 225. 

17. D. Fairlie, T. Ueno : `` Higher Dimensional Generalizations of the Euler 

Top Equation `` hep-th/9710079.

T. Ueno : `` General Solution to the $7D$ Top Equation `` hep-th/9801079.

18. E. Kiritsis : `` Introduction to Non-Perturbative String Theory `` 

hep-th/9708130. 

19. U. Marquard, M. Scholl : Physics Letters {\bf B 227} (1989) 227. 

U. Marquard, R. Kaiser, M. Scholl : Physics Letters {\bf B 227} (1989) 234. 

20. D. Lowe : `` Eleven-Dimensional Lorentz Symmetry from SUSY Quantum 

Mechanics `` hep-th/9807229. 

21. H. Nicolai, R. Helling : `` Supermembranes and M(atrix) Theory ``

hep-th/9809103.

22. S. Melosch : Diploma Thesis. Hambur University (1990),  

23. K. Ezawa, Y. Matsuo and K. Murakami : `` Lorentz Symmetry of Supermembrane 

in the Light Cone Gauge Formulation `` hep-th/9705005.

24. B. de Wit, U. Marquard and H. Nicolai : Comm. Math. Phys {\bf 128} (1990)

39-62

25. C. Castro : Jour. of Chaos, Solitons and Fractals {\bf 7} (7) (1996) 711.

hep-th/9612160  

26-P.S.Howe, O. Raetzel, E. Sezgin : " On Brane Actions and Superembeddings " hep-th/9804051. 

27-S. Ansoldi, C. Castro, E. Spalucci : `` A QCD Membrane `` to appear in Class. Quant. Gravity ``hep-th/0106028

28-. B. Brinne, S. Hjelmeland and U. Lindstrom : " World Volume Supersymmetric Born Infeld Actions " 

hep-th/9904175.

29-L. Marleau : `` Non Abelian Born-Infeld Skyrmions `` hep-th/9408403.

30-E. Guendelman : ``Superextendons witha modified measure  `` hep-th/0006079.

31-S. Ansoldi, C. Castro, E. Spalucci : `` On the Quantum Equivalence 
between Diverse $p$-brane actions ``  Class. Quant. Gravity 
{\bf 17} (2000) L 97  hep-th/0005132.

\bye